\documentclass[preprint,10pt]{JHEP3} 

\JHEPspecialurl{http://jhep.sissa.it/JOURNAL/JHEP3.tar.gz}

\usepackage{epsfig,multicol,amsmath}

\usepackage{epstopdf}

\def\beq{\begin{equation}}
\def\eeq{\end{equation}}
\def\bea{\begin{eqnarray}}
\def\eea{\end{eqnarray}}

\def\eq#1{{Eq.~(\ref{#1})}}
\def\fig#1{{Fig.~\ref{#1}}}
\newcommand{\bas}{\bar{\alpha}_S}

\newcommand{\Lb}{\left(}
\newcommand{\Rb}{\right)}

\parskip=\itemsep               

\renewcommand{\theequation}{\thesection.\arabic{equation}}

\def\thefootnote{\fnsymbol{footnote}}
\def\blfootnote{\xdef\@thefnmark{}\@footnotetext}

\title{\huge \bf Inclusive gluon production in the dipole approach:  AGK cutting rules.}
\author{\Large  Eugene~Levin   \thanks{ leving@post.tau.ac.il,
levin@mail.desy.de;} \,\,and\,\,Alex~Prygarin
 \hspace{0.001cm}
\thanks{ prygarin@post.tau.ac.il
}\\
Department of Particle Physics, School of Physics and Astronomy\\
Raymond and Beverly Sackler
 Faculty
of Exact Science\\  Tel Aviv University, Tel Aviv, 69978, Israel
}

\abstract{
We consider single gluon production in the dipole model  and  reproduce 
the
result of Kovchegov and Tuchin for the adjoint~(gluonic) dipole 
structure of the inclusive cross section.  We show the validity of the 
adjoint dipole structure
to any order of evolution deriving and solving the non-linear 
evolution for the
non-diagonal cross section of a dipole scattering off the target. The 
form of 
the solution to this equation  restores the dipole interpretation for
non-diagonal cross sections, that appear in gluon production. Using this
formalism, we analyse the single inclusive production cross 
section
in terms of the contributions of  different multiplicities, and we 
derive the 
Abramovskii-Gribov-Kancheli(AGK) cutting rules  for two Pomeron 
exchange.
The cutting rules, which were found in this formalism, fully reproduce 
the original AGK
rules for the total cross section. However, for the case of gluon 
production, the  AGK rules are 
violated already for one gluon emission from the vertex. }


\preprint{  TAUP-2876/08\\
 \today}

\begin{document}

\def\thefootnote{\arabic{footnote}}
\section{Introduction}\label{sec:Int}

The high energy scattering in the perturbative QCD has been extensively studied during the last two decades. Special interest is drawn to the situation where the scattering amplitude unitarizes, which is called a saturation regime\cite{GLR,MUQI,MV}. In this regime the density of partons becomes so high that one cannot ignore their mutual interactions anymore. A very convenient framework for studying parton saturation is the so-called dipole approach \cite{MUCD}, where the size of the elementary
degree of freedom, namely, dipole is assumed to be much smaller than $1/\Lambda_{QCD}$ justifying the use of perturbative QCD. A lot of observables were calculated using this extremely convenient formalism. In the present paper we consider the Abramovsky-Gribov-Kancheli~(AGK)~\cite{Abramovsky:1973fm} cutting rules in the framework of the
dipole approach. The AGK cutting rules allow us to expand the dipole approach for the calculation of the scattering elastic amplitude (total cross section) to the consideration of  exclusive processes such as diffractive production \cite{KLDD} and different correlations in multiparticle production processes. They lead also to a better understanding of the $k_t$-factorization in the region of low $x_{Bjorken}$ where the new momentum scale: saturation momentum, makes this factorization questionable.  Being such an important tool, the AGK cutting rules have been
discussed in QCD (see Refs. \cite{AGKQCD}), but the  recent study of the inclusive cross section in QCD has drawn a new attention to these rules. The project was  started by Kovchegov  \cite{Kovchegov:2001ni} and Kovchegov and Tuchin \cite{Kovchegov:2001sc} for the single inclusive cross section and it was   extended to the case of the double gluon production by Kovchegov and Jalilian-Marian \cite{JalilianMarian:2004da}. The main result of this study could be summarized as follows: for the single inclusive cross sections the $k_t$-factorization  has been proved
which leads to a plausible validity of the AGK cutting rules while in the case of the double inclusive cross section the explicit violation of the AGK cutting rules has been found.

The result of Refs.~\cite{Kovchegov:2001ni,Kovchegov:2001sc} for the single inclusive cross sections was confirmed afterwards by Braun \cite{Braun:2006wj} using the reggeized gluon technique, as well as, by Marquet \cite{Marquet:2004xa} and Kovner and Lublinsky  \cite{Kovner:2006wr} in the Wilson lines formalism.

 The result of Ref. \cite{JalilianMarian:2004da} was strongly questioned by Braun \cite{Braun:2006wj}, who claimed that one of the gluons~(upper) is necessarily emitted from the vertex and such a contribution cannot represent a genuine violation of the AGK
cutting rules, because the original
derivation of the AGK rules was based on the assumption that there are no emissions from vertices.

Motivated by this discrepancy in the results, we revisited first   the single inclusive case and found that the result of Kovchegov and Tuchin is correct. In Section~\ref{sec:real-virt} we discuss the key problem  of real-virtual cancellation which is the essential question for approaching the AGK cutting rules problem in QCD. Indeed, the key question for everybody and the nightmare of everybody who is interested in the AGK cutting rules is whether
the set of diagrams that contribute to the total cross section, and to the inelastic production is the same. In this section as well as in the sections~\ref{sec:real-virt}-\ref{sec:evol} we complete the analysis, started by Chen and Mueller \cite{Chen:1995pa} and by Kovchegov and Tuchin\cite{Kovchegov:2001ni,Kovchegov:2001sc}, of the set of  diagrams that are responsible for the  total and inelastic cross sections.

  In Section~\ref{sec:noevol} and Section~\ref{sec:evol}
we repeat the Kovchegov and Tuchin analysis of the single inclusive production but  introducing a new function $M\Lb j k|i k\Rb$ in which it is easy to separate elastic and inelastic interaction. We obtain a new non-linear evolution equation and solve it. Function $M\Lb j k|i k\Rb$ describes the non-diagonal process of scattering of dipole with size $r_{j k}= | x_j - x_k|$ off the target with transition of this dipole to the dipole of the size $r_{i k}=
|x_i - x_k|$. At first sight such an amplitude contradicts the key idea of the dipole approach: dipoles are correct degrees of freedom at high energy or, in other words, the interaction matrix is diagonal with respect to the sizes of the interacting dipoles.  We show by solving the evolution equation for  $M\Lb j k|i k\Rb$, that the dipole approach survives this test of existence the non-diagonal amplitudes at high energy.

Starting from Section~\ref{sec:agk}  we  consider the AGK cutting rules. The AGK cutting rules give us the relation between the total cross section at high energy and the processes of multiparticle production. The main idea stems from the unitarity constraint for the BFKL Pomeron \cite{BFKL} which describes the high energy scattering amplitude in the leading log $(1/x_{Bjorken})$ approximation of perturbative QCD. The unitarity  reads as
\beq \label{UNC}
2 N\Lb Y; x,y\Rb \,\,\,=\,\,| N\Lb Y; x,y\Rb|^2 + G_{in}\Lb Y; x,y\Rb
\eeq
where $Y= (1/x)$ and $(x,y)$ are the coordinates of the incoming dipole; $ N\Lb Y; x,y\Rb$ is the imaginary part of the  elastic amplitude and the first term describes the elastic scattering (assuming that the real part of the amplitude is small at high energy) while the second stands for the contribution of all inelastic processes. In the leading log $(1/x_{Bjorken})$ approximation the elastic contribution can be neglected and for the BFKL Pomeron \eq{UNC} can be reduced to the form (see \fig{cutpom}):
\beq \label{UNPO}
2 N^{BFKL}\Lb Y; x,y\Rb \,\,=\,\, G^{BFKL}_{in}\Lb Y; x,y\Rb
\eeq
In  what follows we call $G_{in}\Lb Y; x,y\Rb$ a cut Pomeron while $N^{BFKL}\Lb Y; x,y\Rb$ will be called a Pomeron or uncut Pomeron.
\FIGURE[ht]{\begin{minipage}{85mm}
\centerline{\epsfig{file=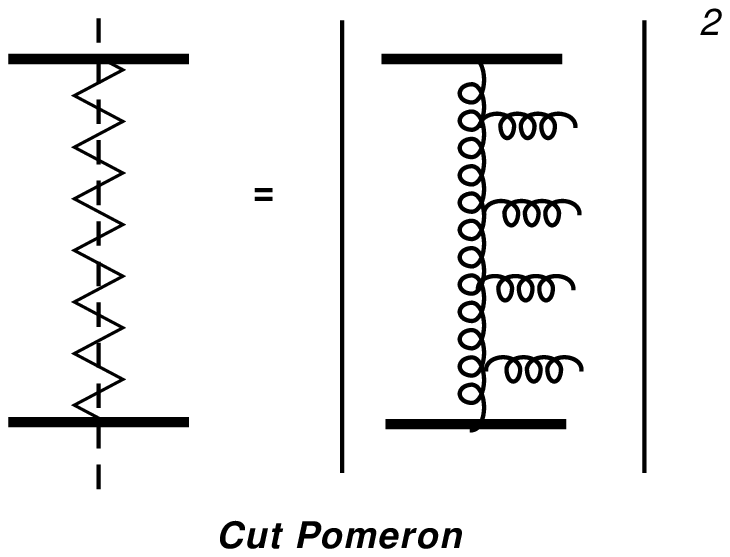,width=80mm,height=40mm}}
\end{minipage}
\caption{The definition   of cut Pomeron through the BFKL ladder. }
\label{cutpom} }

In this paper we prove  the AGK cutting rules for the case of the totally inclusive processes (see \fig{agkin}). We show that the same set of the diagrams determine the total cross section and the processes of the inelastic production. Therefore, each   inelastic contribution for the exchange of two BFKL Pomerons can be calculated in terms of the elastic amplitude with the coefficients shown in \fig{agkin}.
The situation changes drastically for the case when we measure one extra particle (see \fig{agkver}.
In this case the inelastic production stems from  different set of the diagrams than the total cross section and we do not have the simple expression for each inelastic processes through the elastic amplitude as well as we do not find the simple relations between different processes.

\FIGURE[h]{ \centerline{\epsfig{file= 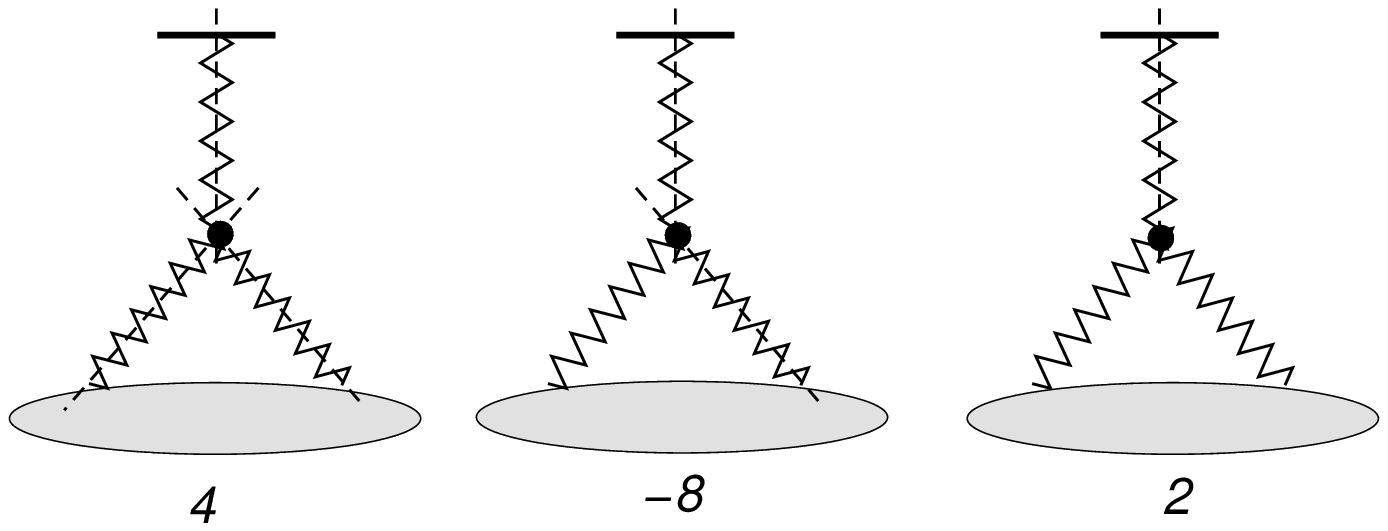,width=120mm,height=35mm }}
\caption{ The AGK cutting rules for  the total inclusive processes. Cut Pomeron is defined in \protect\fig{cutpom} and in \protect\eq{UNPO}.
} \label{agkin} }
\FIGURE[h]{ \centerline{\epsfig{file= 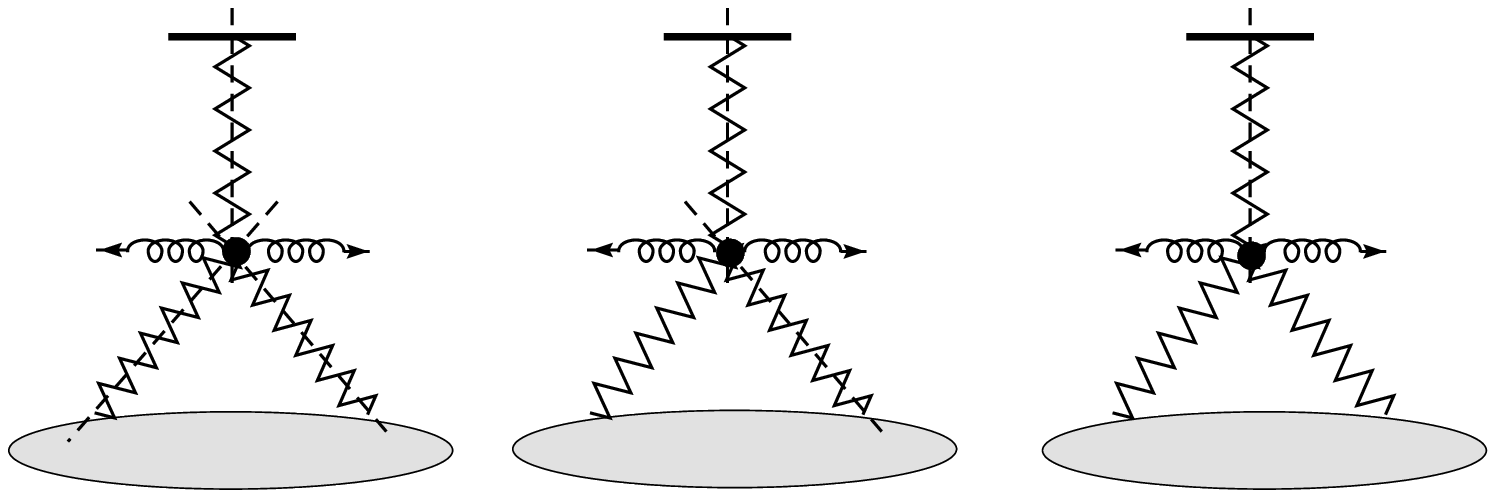,width=120mm,height=35mm}}
\caption{ Violation of AGK cutting rules for the processes with additional emission of one particle (gluon). Cut Pomeron is defined in \protect\fig{cutpom} and in \protect\eq{UNPO}.
} \label{agkver} }

In the Appendices we adduce some calculations relevant to our discussion.

\section{Real-virtual cancellations}\label{sec:real-virt}
In this section we briefly review the main result of  Ref.~\cite{Chen:1995pa} with a special emphasis on so-called real-virtual cancellations. These cancellations play an important role in  the  derivation of both BFKL~\cite{BFKL} and BK~\cite{B,K} equations in the colour dipole model. Let us consider a colourless onium state with one extra soft gluon emission.
The soft gluon can be emitted either from a quark or from a antiquark lines. For simplicity we consider here only the emissions from the antiquark line. We assign the transverse coordinates $x_1$, $x_0$ and $x_2$ to the quark, antiquark and soft gluon lines, respectively. The system interacts with the target by instantaneous interaction via
Coulomb gluon exchange. We use an eikonal approximation allowing for multiple interaction with the target. It was shown by Kovchegov \cite{Kovchegov:2001ni} that there are no soft gluon emissions during the interaction time so that the eikonal
rescattering can be regarded as an instantaneous as well. To avoid any question regarding the use of  the eikonal approximation as the initial condition for the evolution at low energy, we consider the interaction with heavy nuclei for which the eikonal formula, as well as, the non-linear evolution equation in the mean field approximation (Balitsky-Kovchegov (BK) equation \cite{B,K}) are proved.  We work in Light Cone Perturbation Theory~(LCPT)\cite{BRLE} with light-cone gauge and denote by $\tau=0$  the interaction time while the detector that measures the particles in the final state is placed at $\tau=\infty$.

For simplicity, we consider only soft gluon emissions from the antiquark with coordinate $x_0$.
The relevant diagrams are shown in   \fig{fig:real-virt}

\FIGURE[ht]{ \centerline{
\begin{tabular}{ccc}
\epsfig{file=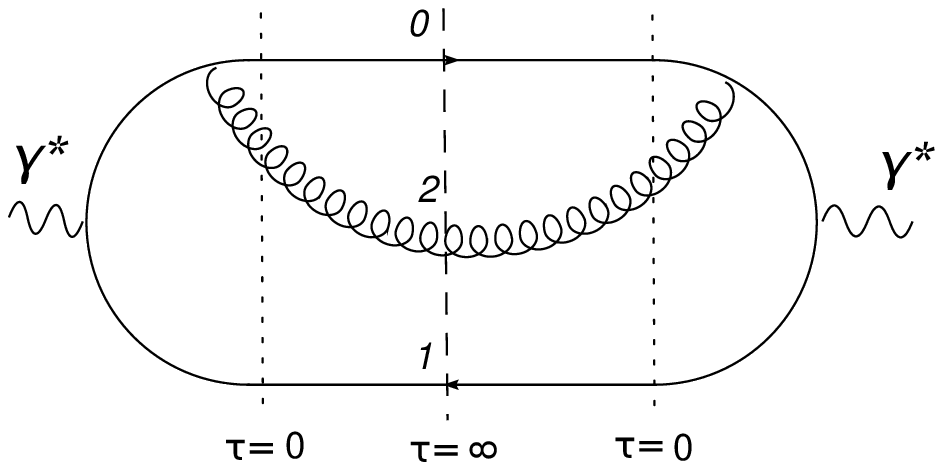,width=50mm} & \epsfig{file=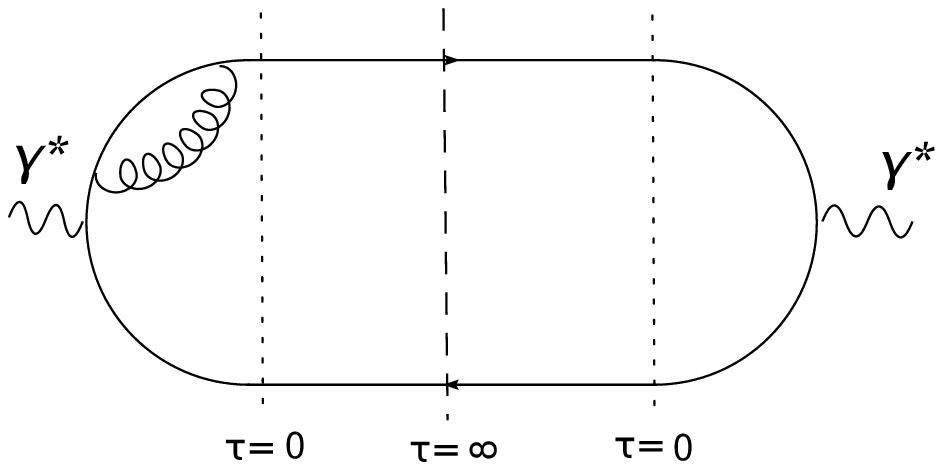,width=50mm}&
\epsfig{file=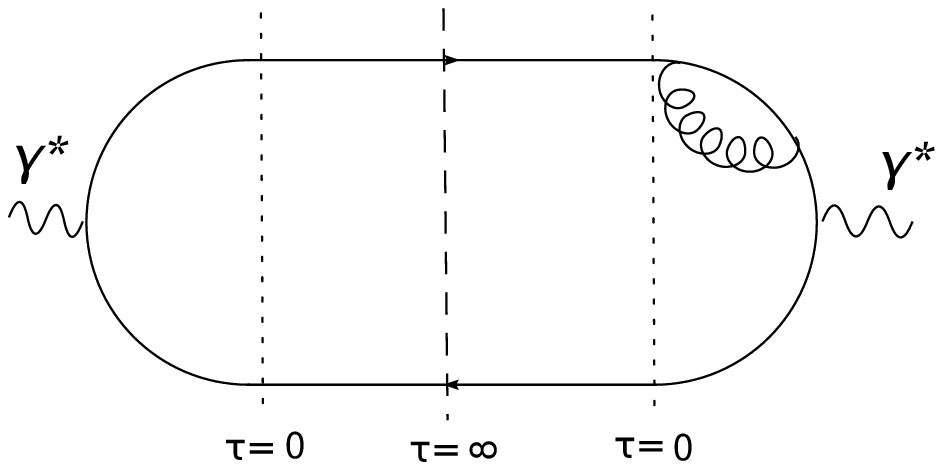,width=50mm}\\
$A$ & $R$ & $R^*$
\\
\epsfig{file=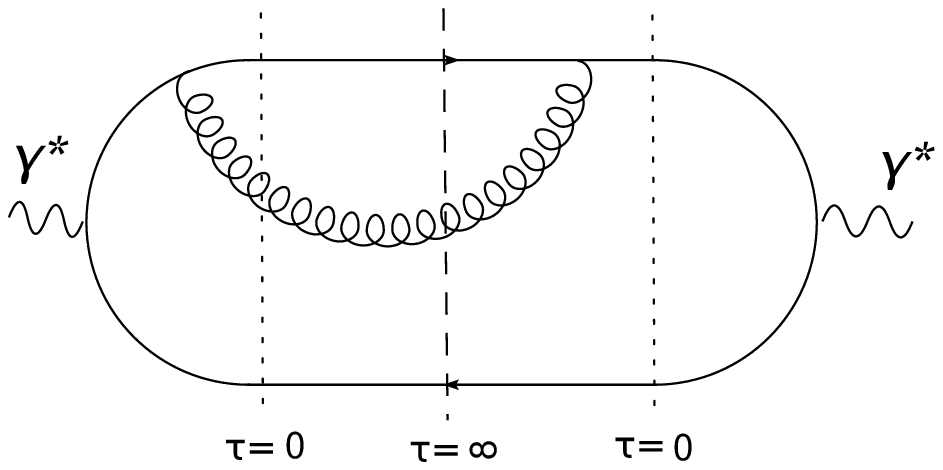,width=50mm} & \epsfig{file=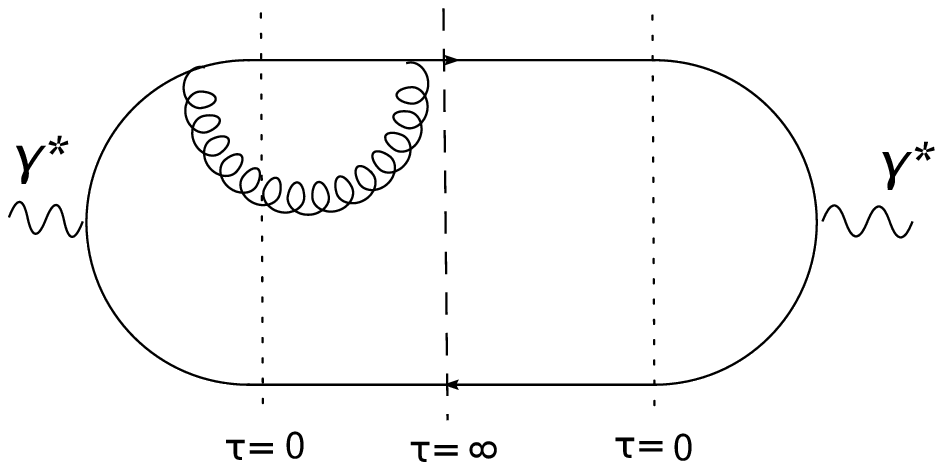,width=50mm}&
 \\
$B$ & $C$ & \\
\epsfig{file=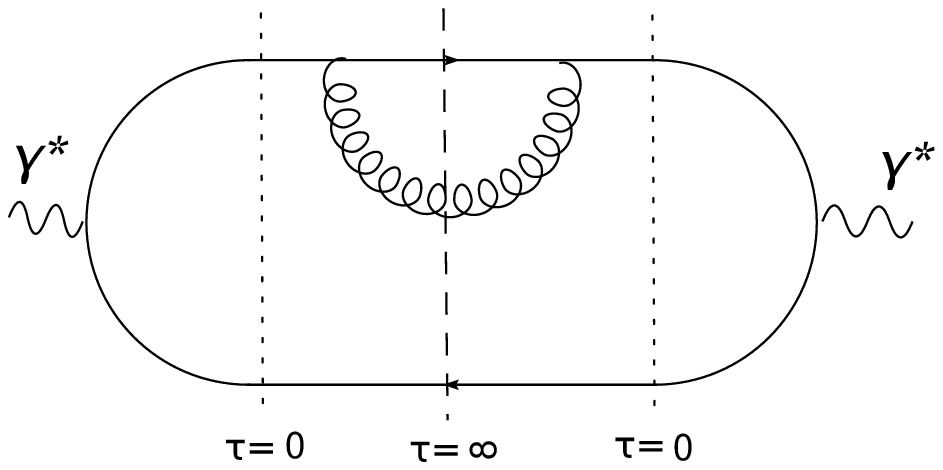,width=50mm} & \epsfig{file=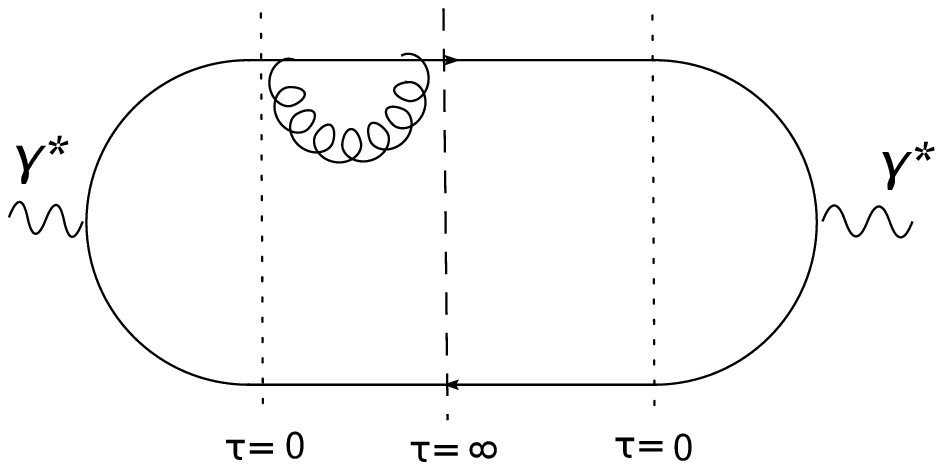,width=50mm}&
\epsfig{file=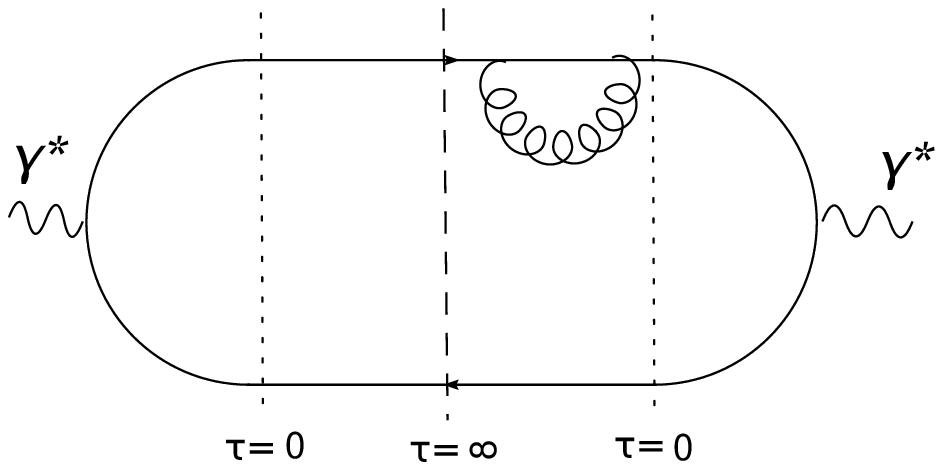,width=50mm}\\
$D$ & $E$ & $F$
\end{tabular}
} \caption{All possible emissions of a soft gluon in the onium state, diagrams $B^*$ and $C^*$ are not
shown.}\label{fig:real-virt} }


In diagram $A$ of \fig{fig:real-virt} the soft gluon at $x_2$ is emitted before the interaction time in both amplitude and conjugate amplitude. In diagrams $R$ and $R^*$ the soft gluon is emitted and absorbed before the interaction time, these diagrams give reggeization term of the BFKL and BK equations in the dipole model. Only diagrams $A$, $R$ and $R^*$ are present in the BFKL\cite{BFKL} and BK \cite{B,K} equations, all other diagrams are canceled as we will show shortly. This cancellation is called real-virtual cancellation, where any real emission~(absorption) of the soft gluon after time $\tau=0$ is canceled by its virtual counterpart(s). Namely, let us consider diagrams $B$ and $C$ where the soft gluon is emitted before the interaction time in the amplitude, but not in the conjugate amplitude. The emitted gluon be either present or not in the final state at $\tau=\infty$. As it was shown in Ref.~\cite{Chen:1995pa}~( see also Appendix~A for more details) those diagrams differ only by a minus sign and thus cancel each other in the total cross section (same for $B^*$ and $C^*$, not shown here). Similar cancellation happens also to $D$, $E$ and $F$, but one should note that $E$ and $F$ have a factor of $\frac{1}{2}$ w.r.t. diagram $D$ due to a light-cone time ordered integral as shown in Appendix of Ref.~\cite{Chen:1995pa}.

Thus we are left with diagrams $A$, $R$ and $R^*$. To this we should add  also diagrams with the soft gluon emission from 
the quark line at $x_1$, which translates into the BFKL kernel in the dipole model, namely, $\left(\frac{x_{12}}{x^2_{12}}-\frac{x_{02}}{x^2_{02}}\right)^2=\frac{x^2_{10}}{x^2_{12}x^2_{20}}$. 

In Section~\ref{sec:evol} we present the derivation of BK equation in more general case and show how this reduces to
well-known equation 
\begin{eqnarray}\label{BK}
\frac{\partial N(10)}{\partial y}=\frac{\bar{\alpha_s}}{2\pi}\int d^2x_2\frac{x^2_{10}}{x^2_{12}x^2_{20}}\left\{N(12)+N(02)-N(10)-N(12)N(02)\right\}
\end{eqnarray}
with $N(10) = \mbox{Im} A_{el}(x_{1},x_{0};Y)$ being  the imaginary part  \footnote{   For simplicity, we omit the argument $Y = log (1/x_{Bjorken})$ where $x_{Bjorken}$ is the fraction of energy carried by the dipole.} of  the elastic scattering amplitude of a colourless dipole with quark and antiquark coordinates at $x_1$ and $x_0$ ($\vec{x}_{10} = \vec{x}_1 - \vec{x}_0$), respectively.

\section{Inclusive one gluon production: no evolution included}\label{sec:noevol}
In this section we follow the lines of Ref.~\cite{Kovchegov:2001ni} in the derivation of the single gluon inclusive production cross
section  in DIS with no evolution included. However, in our derivation we treat separately all the contributions that sum into the final result of Ref.~\cite{Kovchegov:2001ni}. The reason for that is our goal  to  include properly the evolution in the inclusive cross section for one gluon production as will be clarified in Section~\ref{sec:evol}.

The general expression for the gluon production cross section in DIS is given by  
\begin{eqnarray}\label{sigmaDIS}
\frac{d\sigma^{\gamma^* A \rightarrow GX}}{d^2k\;dy}=\frac{1}{2\pi^2}\int d^2x_{10}\;dz |\psi^{\gamma^* \rightarrow q\bar{q}}(x_{01},z)|^2 
\frac{d\sigma^{q\bar{q} A \rightarrow GX}(x_1,x_0) }{d^2k\;dy}
\end{eqnarray}
where $d\sigma^{q\bar{q} A}(x_1,x_0) /d^2k\;dy$ is the gluon production cross section 
for the scattering of a colour dipole with quark coordinate $x_1$ and antiquark coordinate $x_0$ on the target, and 
$\psi^{\gamma^* \rightarrow q\bar{q}}(x_{01},z)$ is the well-known \cite{barone} wave function of the splitting of the virtual photon in DIS into $q\bar{q}$ pair with of a transverse size $x_{10}=x_1-x_0$ and a fraction $z$ of the longitudinal momentum of the photon carried by the quark.

We want to point out that we limit our discussion to a case where the produced gluon is the hardest gluon emitted
in $q\bar{q}$ system. 

We start with selecting only those diagrams from \fig{fig:real-virt}, which have a real gluon emission ( the soft gluon appears at the cut at $\tau=\infty$). It is easy to see that we are left with $A$, $B$ ($B^*$) and $D$.
In the inclusive cross section the transverse momentum $k$ of the soft gluon is kept fixed and thus transverse coordinates of the produced gluon are different in the amplitude ($x_2$) and the conjugate amplitude ($x_{2'}$). 
Before writing the expression for the inclusive cross section with one gluon production   we define an object which will 
play a central role in  our further derivations. The function $M(12|34)$ is defined as an unintegrated over the impact parameter cross section of the scattering of a dipole with coordinates of quark (antiquark) being $x_1$ ($x_2$) in the amplitude and  $x_3$ ($x_4$) in the conjugate amplitude. One should keep in mind that the cross section $M(12|34)$ is a function of rapidity though it is not reflected in our notation.
 The explicit expression for $M(12|34) = M_0(12|34)$ in the case of interaction with a nucleus without evolution is given in Appendix~\ref{sec:B}.
  In fact, we have three of them, for each of the diagrams $A$, $B$ ($B^*$) and $D$ in \fig{fig:real-virt}.
 In Section~\ref{sec:evol} we show that all three functions are described by the same evolution equation and obey a generalized form of the optical theorem for the case of the dipole having different sizes in the amplitude and the conjugate amplitude.

We are  ready to  write the expression for the inclusive cross section in terms of $M^A$, $M^B$ and $M^D$ as follows \footnote{In \eq{line1} and everywhere in this paper we use notation $k\,x$ instead of $\vec{k}_{\perp} \cdot \vec{x}_{\perp}$. We hope that this will not cause difficulty in understanding.}

 \footnotesize
\begin{eqnarray}\label{line1}
\frac{d\sigma^{q\bar{q} A \rightarrow GX}(x_1,x_0)}{d^2kdy}=\frac{\bar{\alpha_s}}{2\pi}\frac{1}{(2\pi)^2}\int d^2x_2 d^2{x}_{2'}e^{-ik(x_2-x_{2'})}\left\{\left(\frac{x_{12}}{x^2_{12}}-\frac{x_{02}}{x^2_{02}}\right)\left(\frac{x_{12'}}{x^2_{12'}}-\frac{x_{02'}}{x^2_{02'}}\right)\left(M_0(12|12')+M_0(20|2'0) 
\right. \right.
\end{eqnarray} 
\begin{eqnarray}\label{line2}
\left. \left.
+M_0(12|12')M_0(20|2'0)-M_0(12|12')\left\{N_0(20)+N_0(2'0)\right\}-M_0(20|2'0)\left\{N_0(12)+N_0(12')\right\}+N_0(20)N_0(12')+N_0(12)N_0(20')\right)
\right.  \;\;\;\;\;\; 
\end{eqnarray}
\begin{eqnarray}\label{line3}
\left.
-\left(\frac{x_{12}}{x^2_{12}}-\frac{x_{02}}{x^2_{02}}\right)\left(0-\frac{x_{02'}}{x^2_{02'}}\right)
\left\{M_0(12|10)\left(1-N_0(20)\right)+N_0(20)N_0(10)\right\}
\right.
\end{eqnarray}
\begin{eqnarray}
\left.
-\left(\frac{x_{12}}{x^2_{12}}-\frac{x_{02}}{x^2_{02}}\right)\left(\frac{x_{12'}}{x^2_{12'}}-0\right)
\left\{
M_0(20|10)\left(1-N_0(12)\right)+N_0(10)N_0(12)\right\} \hspace{0.5cm}
\right.  \nonumber
\end{eqnarray}
\begin{eqnarray}\label{line4}
\left.
-\left(0-\frac{x_{02}}{x^2_{02}}\right)\left(\frac{x_{12'}}{x^2_{12'}}-\frac{x_{02'}}{x^2_{02'}}\right)
\left\{
M_0(10|12')\left(1-N_0(2'0)\right)+N_0(10)N_0(2'0)\right\}
\right.
\end{eqnarray}
\begin{eqnarray}
\left.
-\left(\frac{x_{12}}{x^2_{12}}-0\right)\left(\frac{x_{12'}}{x^2_{12'}}-\frac{x_{02'}}{x^2_{02'}}\right)
\left\{M_0(10|2'0)\left(1-N_0(12')\right)+N_0(10)N_0(12')\right\}
\right. \nonumber
\end{eqnarray}
\begin{eqnarray}\label{line5}
+\left(\frac{x_{12}}{x^2_{12}}\frac{x_{12'}}{x^2_{12'}}+\frac{x_{02}}{x^2_{02}}\frac{x_{02'}}{x^2_{02'}}\right)M_0(10|10)
-\left(\frac{x_{12}}{x^2_{12}}\frac{x_{02'}}{x^2_{02'}}+\frac{x_{02}}{x^2_{02}}\frac{x_{12'}}{x^2_{12'}}\right)N^2_0(10)
\left.
\right\}
\end{eqnarray} 
\normalsize
where $N_0$ is the initial condition of BK  equation and  $M_0(ik|il)$ is given in Appendix~\ref{sec:B}. 
 
In \eq{line1} we account for a separate rescattering of dipoles $"12"("12'")$ and $"20"("2'0")$ coming from diagram~$A$. 

In \eq{line2} we include the contributions where both of the dipoles are rescattered at the same time in diagram $A$. The first term describes the case where  the two dipoles are rescattered both elastically and inelastically. The second and the third terms represent both elastic and inelastic rescattering of one dipole and only elastic (either amplitude or conjugate amplitude) rescattering of the other dipole. The fourth and fifth terms describe 
only elastic rescattering of the two dipoles. One should note that terms  $N_0(20)N_0(12')+N_0(12)N_0(20')$ are not 
included into  $M_0(12|12')M_0(20|2'0)$ since the elastic part of the  cross section $M_0(12|12')$ by the definition starts with the lowest order of $N_0(12)N_0(12')$ (there are no disconnected diagrams), and thus the terms $N_0(20)N_0(12')+N_0(12)N_0(20')$ should be added separately.

In \eq{line3} we have contributions from diagram $B$($B^*$) which are not symmetrical in the Kernel. Namely, some diagrams are prohibited by the definition of the cross section $M_0(12|10)$. As a simple example, we try to include contributions which do not enter \eq{line3}. We consider
 $\frac{x_{12}}{x_{12}^2}\frac{x_{12'}}{x_{12'}^2}M_0(12|10)$ as shown in \fig{fig:B-Example}. It is immediately
 clear that $M_0(12|10)$ cannot multiply such a dipole splitting, since  a dipole $"12"("10")$ is not present at the interaction time  $\tau=0$ in the conjugate amplitude with such a dipole splitting $\frac{x_{12}}{x_{12}^2}\frac{x_{12'}}{x_{12'}^2}$. In other words, the dipole formed by the lower quark loop scatters only elastically.

\FIGURE[h]{ \centerline{
 \epsfig{file=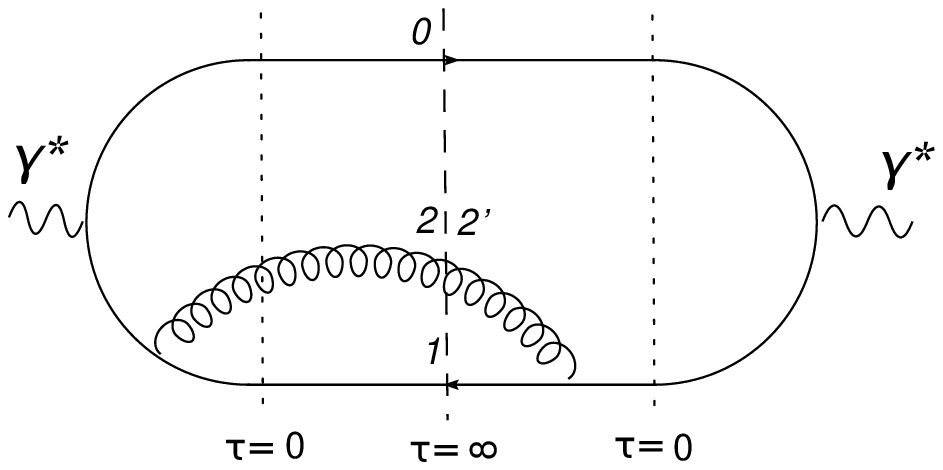,width=50mm} } \caption{ An example of a diagram that leads to asymmetric Kernels in \eq{line3} and \eq{line4}. }\label{fig:B-Example}}

The first term in  \eq{line3} includes elastic and inelastic rescattering of only one dipole, and also 
elastic rescattering of the other dipole. The term $N_0(20)N_0(10)$ corresponds to the elastic scattering of the dipole $"20"$ in the amplitude and the elastic scattering of the dipole 
$"10"$ in the conjugate amplitude, as it was mentioned before this term is not present in $M(20|10)N(20)$ 
by the definition of $M(20|10)$ due to the absence of the disconnected diagrams.
Same for all other terms in
\eq{line3}  and \eq{line4}.  

Finally, in \eq{line5} we account for the contribution from diagram $D$, where we have both elastic and inelastic scatterings of the initial dipole $"10"$. This can be easily understood from all $D$-type diagrams, where the measured 
gluon is emitted from quark and/or antiquark after the interaction time $\tau=0$. In the case of "crossed" dipole splittings $\frac{x_{12}}{x^2_{12}}\frac{x_{02'}}{x^2_{02'}}$ etc. the inelastic rescattering is suppressed in  large $N_c$, that is the reason why the last term in \eq{line5} has $N^2_0(10)$ instead of $M(10|10)$(which 
also includes the inelastic part). 

As the last step in our derivation we have to substitute the expressions for the 
total cross section $M_0$ and  the elastic amplitude $N_0$ by their eikonal formulae
which are found in Appendix~\ref{sec:B}.  After some tedious, but straightforward algebra we obtain

\begin{eqnarray}\label{Classline1}
\frac{d\sigma^{q\bar{q} A \rightarrow GX}(x_1,x_0)}{d^2kdy}=\frac{\bar{\alpha_s}}{2\pi}\frac{1}{(2\pi)^2}\int d^2x_2 d^2{x}_{2'}e^{-ik(x_2-x_{2'})}\left\{ \frac{x_{02}}{x^2_{02}}\frac{x_{02'}}{x^2_{02'}}
\left(1+e^{-2x^2_{22'}Q^2_s/4} -e^{-2x^2_{20}Q^2_s/4}-e^{-2x^2_{2'0}Q^2_s/4}\right)
\right.  \hspace{1cm}
\end{eqnarray} 
\begin{eqnarray}\label{Classline2}
\left. - \frac{x_{12}}{x^2_{12}}\frac{x_{02'}}{x^2_{02'}}
\left(e^{-2x^2_{10}Q^2_s/4}+e^{-2x^2_{22'}Q^2_s/4} -e^{-2x^2_{12'}Q^2_s/4}-e^{-2x^2_{20}Q^2_s/4}\right)
\right.  \;\;\;\;\;\; 
\end{eqnarray}
\begin{eqnarray}\label{Classline3}
\left.
- \frac{x_{02}}{x^2_{02}}\frac{x_{12'}}{x^2_{12'}}
\left(e^{-2x^2_{10}Q^2_s/4}+e^{-2x^2_{22'}Q^2_s/4} -e^{-2x^2_{02'}Q^2_s/4}-e^{-2x^2_{21}Q^2_s/4}\right)
\right. 
\end{eqnarray}
\begin{eqnarray}\label{Classline4}
\left.
+\frac{x_{12}}{x^2_{12}}\frac{x_{12'}}{x^2_{12'}}
\left(1+e^{-2x^2_{22'}Q^2_s/4} -e^{-2x^2_{12}Q^2_s/4}-e^{-2x^2_{12'}Q^2_s/4}\right)
\right\}
\end{eqnarray} 
\normalsize

which is the result obtained by Kovchegov in \cite{Kovchegov:2001ni}. At first sight, there is a difference of the factor of $2$ in the exponentials, which comes from effective double interaction of the same dipole. At this point we have to clarify this coefficient in powers of the exponentials in Eqs.~(\ref{Classline1})-(\ref{Classline4}). We found some confusion in the literature regarding this coefficient, it varies from  $1/4$ to $1/2$ depending on the author and even sometimes for the same author. This confusion stems from the definition of the saturation scale for a quark or a gluon which differs by a factor of $2$. We use the quark saturation scale throughout the paper having a coefficient $1/2$ in the exponentials of Eqs.~(\ref{Classline1})-(\ref{Classline4}). The physical meaning of this is explained as follows.

The important feature of the expression for the inclusive single gluon production  Eqs. (\ref{Classline1})-(\ref{Classline4}) is that interaction happens only to the emitted soft gluon and a quark(antiquark) from which it was emitted. The interaction of a quark~(antiquark) which does not emit the soft gluon cancels and can be schematically explained as follows. In terms of the amplitude, diagrams $A$, $B$($B^*$) and $D$ in \fig{fig:real-virt} have only  two possibilities for the real soft gluon: to emitted either before or after the interaction time $\tau=0$ as shown in \fig{fig:amp}.

\FIGURE[h]{ \centerline{
 \epsfig{file=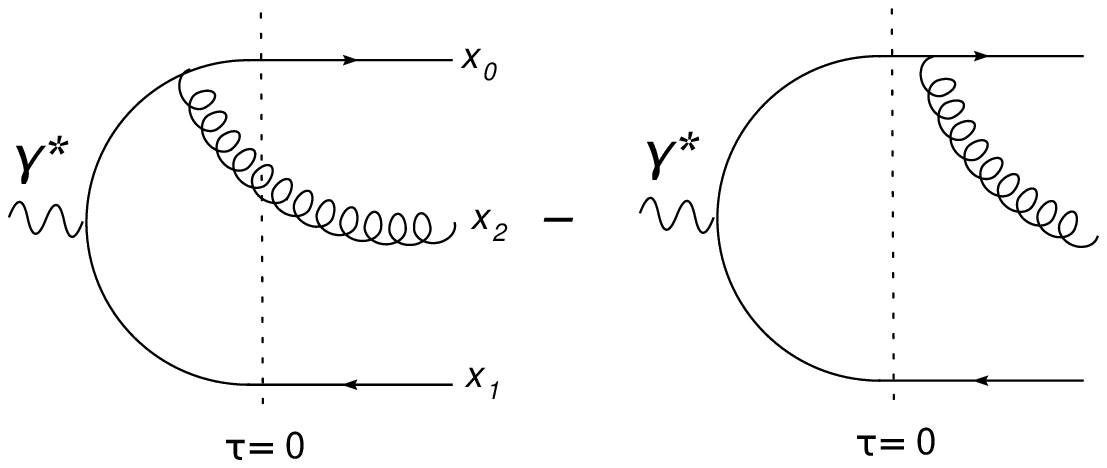,width=70mm} } \caption{ This illustrates the fact that in the low energy limit the interaction occurs only with the produced gluon and a quark~(antiquark) from which it was emitted. The cancellation leads to effective interaction twice with the same (upper) dipole.}\label{fig:amp}}

  The emission after $\tau=0$ has a minus sign w.r.t the emission before $\tau=0$, which can be easily checked by considering light-cone energy denominators (see for example \cite{Chen:1995pa,Kovchegov:2001ni,Kovchegov:2001sc,JalilianMarian:2004da} and also Appendix~A). For simplicity, consider one $t$-channel gluon interaction with the target  for each 
dipole $(12)$ and $(20)$ formed by the soft gluon emission. In this case we have two contributions in the transverse 
coordinate space for the lower dipole
\begin{eqnarray}\label{amp-low}
\int \frac{d^2l}{l^2}\left\{\left(e^{-ilx_1}-e^{-ilx_0}\right)-\left(e^{-ilx_1}-e^{-ilx_2}\right)\right\}=
\int \frac{d^2l}{l^2} \left(e^{-ilx_2}-e^{-ilx_0}\right) 
\end{eqnarray}
the upper dipole has only one contribution
\begin{eqnarray}\label{amp-up}
\int \frac{d^2l}{l^2} \left(e^{-ilx_2}-e^{-ilx_0}\right) 
\end{eqnarray}  
From \eq{amp-low} and \eq{amp-up} one can see that the two contributions for the lower dipole effectively sum into interaction with the upper dipole. It should be mentioned that one cannot sum \eq{amp-low} and \eq{amp-up} in the amplitude since they have different colour matrices, but after multiplication by conjugate amplitude one obtains the same colour factor which results into a factor of $2$ in the exponentials in Eqs. (\ref{Classline1})-(\ref{Classline4}) and looks like a dipole interacts twice with the target~(or two dipoles of the same size interact once). 
In the BK equation (written for total cross section) we do not observe this, since the second diagram in \fig{fig:amp} cancels with corresponding virtual gluon emission.
 This \emph{ad hoc} explanation is by no means an exhaustive one, but rather a simple argument why one should expect 
only the soft gluon and the quark~(antiquark) from which it was emitted to interact with the target. A rigorous calculation shows the same result, provided we assign to the interaction with target the same expression 
 before and after the gluon emission. 
Here we demonstrated how this happens in the case of the  eikonal interaction at low energy
with no evolution included.
In the next Section we show that proper inclusion of the non-linear evolution effects does not change this general feature of the single inclusive cross section that the "spectator quark" never interacts with the target in the course of the evolution and  thus all the interaction can be expressed in terms of the adjoint amplitude $N_G(x)=2N(x)-N^2(x)$ as was suggested by Kovchegov and Tuchin in Ref.~\cite{Kovchegov:2001sc}.


\section{Inclusive gluon production with evolution}\label{sec:evol}

In this section we want to include the evolution into the expression for the single gluon inclusive cross section found in Section \ref{sec:noevol}. 
It was suggested in Ref.~\cite{Kovchegov:2001sc} that each exponential in Eqs. (\ref{Classline1})-(\ref{Classline4}) is to be replaced with a function $1-N_G$, where $N_G$ is the elastic scattering amplitude of an adjoint ("gluonic") dipole
  given by $N_G(22')=2N(22')-N^2(22')$ with $N(22')$ being a solution for the BK equation.This substitution is based on the classical expression where only the produced gluon and a quark(antiquark) from which it was emitted do interact with the target and thus  in the course of the evolution one can safely ignore the spectator quark~(antiquark). 
Kovchegov and Tuchin \cite{Kovchegov:2001sc} analyzed the next step of the evolution in the rapidity where an extra  softer gluon is emitted 
and found that this form also holds to this order. This way the authors  generalized the result to any order of the softer gluon emissions. 
 In this Section we use the expression for the single inclusive cross section Refs.~(\ref{line1}-\ref{line5}) supplemented by the evolution equation for the function $M(ij|ik)$ to show explicitly the validity of the adjoint("gluonic")
dipole structure at any order of the evolution.

We start with deriving the evolution equation for the function $M(ij|ik)$, which has the meaning of the total cross section for $j=k$. It should be emphasized that $M(ij|ik)$ is also a function of the rapidity though it is not indicated in our notation. We are interested in the evolution in rapidity where only softer gluons are emitted, the emission of the harder gluons can be easily included through the dipole density as it was shown by Kovchegov and Tuchin \cite{Kovchegov:2001sc}, and thus is not relevant for our discussion. We want to remind that we are interested only in the situation where the measured gluon is the hardest gluon in the system.
In our derivation we help with a set of the simple mnemonic rules, as follows,
\begin{itemize}
\item  the gluon is represented by a double quark line in the large $N_C$ limit, the triple gluon vertex can be effectively written as an emission of the softer gluon from the quark and antiquark components of the harder gluon.
The emission from the antiquark component of the harder gluon has a relative minus sign w.r.t. the emission from the quark component. For more details, see Ref.~\cite{Chen:1995pa};
\item the virtual emission (diagrams $R$,$C$, $E$ and $F$ in \fig{fig:real-virt} ) has relative minus sign w.r.t. to the real emissions (diagrams $A$, $B$ and $D$ in \fig{fig:real-virt}). In addition to this, the diagrams where the virtual gluon is not present at the interaction time $\tau=0$ has a factor of $1/2$ due to light cone time ordered integration. For more details, see Ref.~\cite{Chen:1995pa};
\item the softer gluon can be emitted only after the harder gluon before the interaction time $\tau=0$, and
only before the harder gluon after the interaction time. In the case where one of the gluons is emitted before $\tau=0$ and the other one after $\tau=0$, the light cone time ordering is arbitrary, irrespectively whether the soft or the hard gluon was emitted before $\tau=0$. These light cone time ordering rules where derived by Jalilian-Marian and Kovchegov  \cite{JalilianMarian:2004da} using light cone energy denominators.    
\end{itemize}

We would like to comment more on the last light cone time ordering rule. This rule implies that a softer gluon can be emitted only from a dipole which was present at $\tau=0$, this means the all the information about further evolution can be encoded in the function $M(ij|ik)$ through specifying the coordinates of the dipoles present at $\tau=0$ in the amplitude and conjugate amplitude (and, of course, the energy variable not reflected in our notation).  In the more general case, where the rapidity separation between two adjacent emitted gluons is not large (multi Regge kinematics does not hold) one has also to specify the dipoles present at $\tau=\infty$.

We are ready to write the evolution equation for the function $M(12|12')$, which has a meaning of the total scattering cross section for a dipole having quark at $x_1$ and antiquark at $x_2$ in the amplitude, as well as, quark at $x_1$ and antiquark at $x_2'$ in the complex conjugate amplitude. We analyze the emission of a softer gluon with transverse coordinate $x_3$ in the dipole $"12"("12'")$.
 Working in the large $N_C$ limit we represent the gluons by double quark-antiquark line and show them disconnected from other lines reflecting all possible couplings to quark or antiquark line in accordance with the notation used in \cite{Kovchegov:2001sc}. 
 In analogy with \fig{fig:real-virt} consider all possible emissions of gluon $"3"$ in the dipole $"12"("12'")$ depicted in \fig{fig:22tag3} ( diagram $B^*$ and $C^*$ are not shown, but must be included as well).

\FIGURE[h]{ \centerline{
\begin{tabular}{ccc}
\epsfig{file=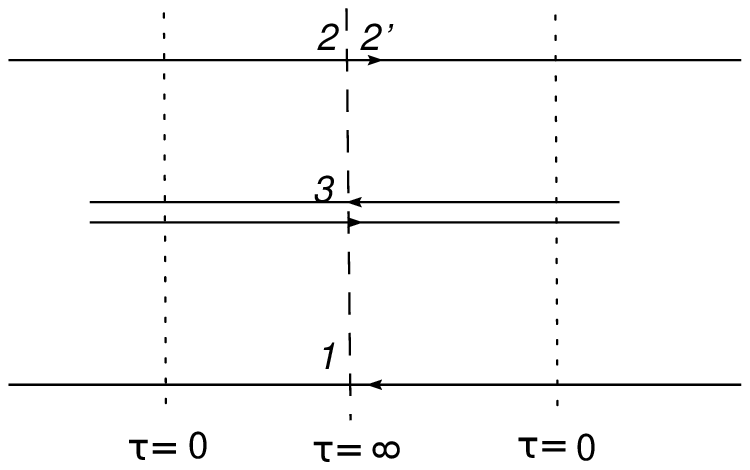,width=40mm} & \epsfig{file=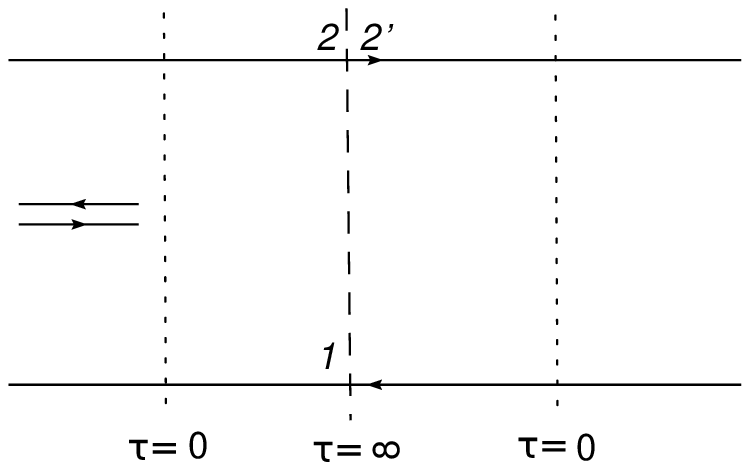,width=40mm}&
\epsfig{file=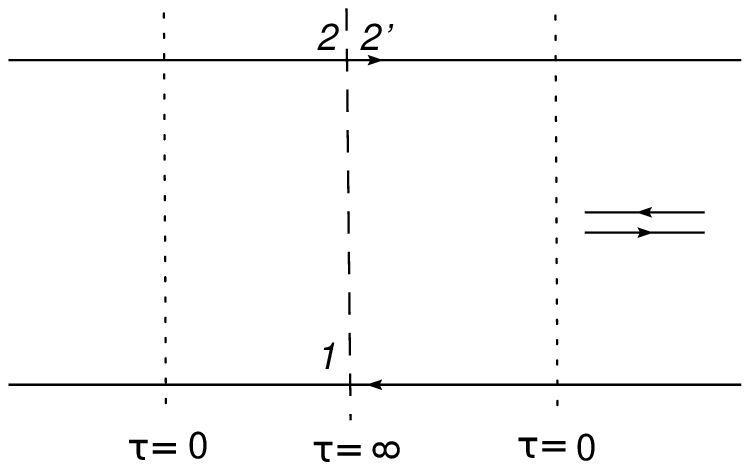,width=40mm}\\
$A$ & $R$ & $R^*$
\\
\epsfig{file=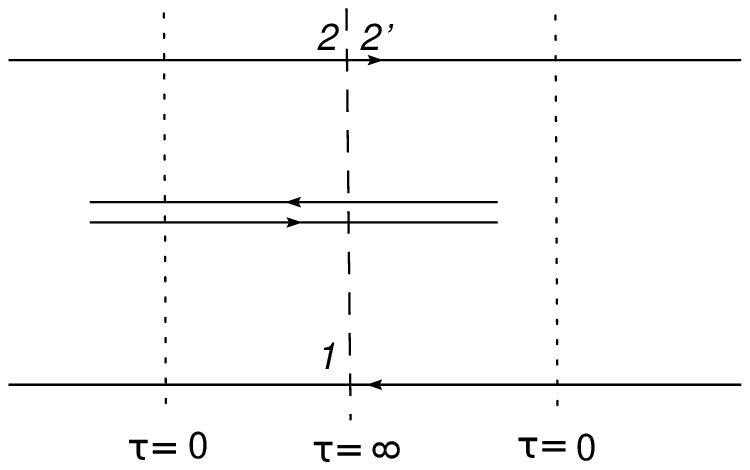,width=40mm} & \epsfig{file=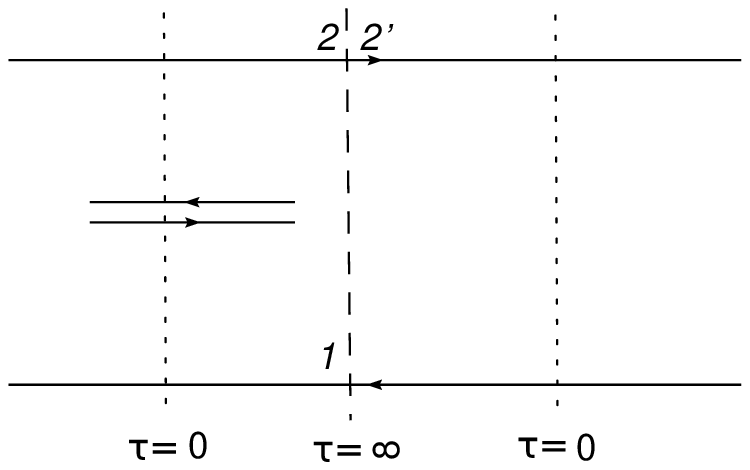,width=40mm}&
 \\
$B$ & $C$ & \\
\epsfig{file=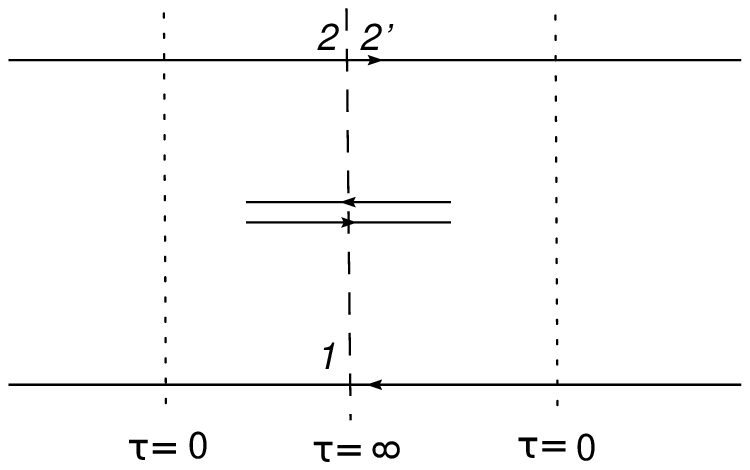,width=40mm} & \epsfig{file=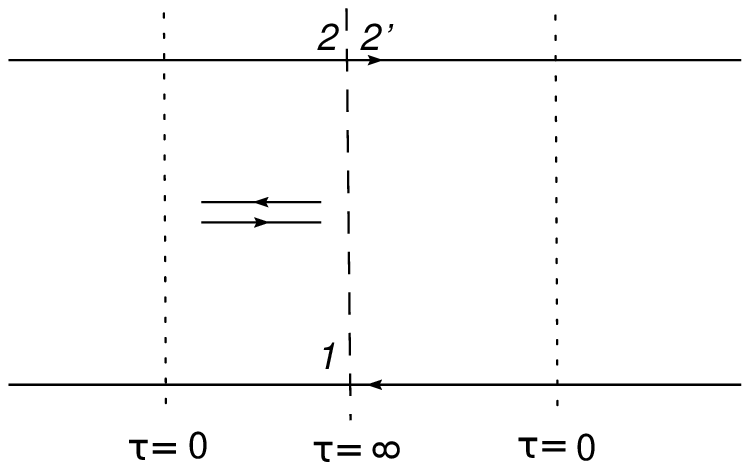,width=40mm}&
\epsfig{file=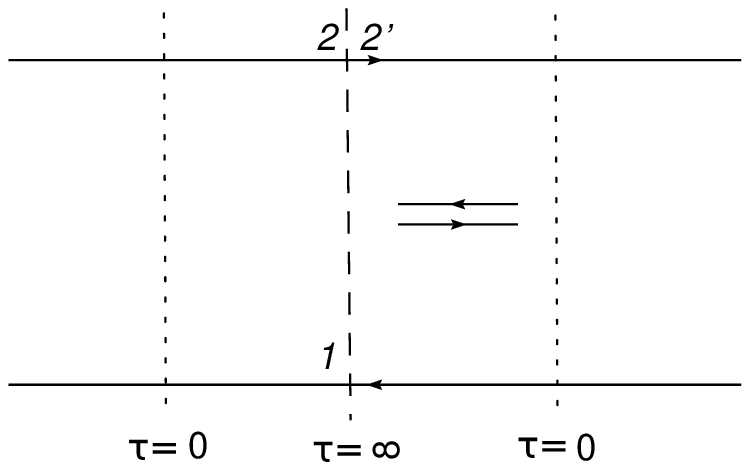,width=40mm}\\
$D$ & $E$ & $F$
\end{tabular}
} \caption{ All possible emissions of the softer gluon $"3"$ in the dipole $"12"("12'")$, diagrams $B^*$ and $C^*$
are not shown. }\label{fig:22tag3} }


The evolution equation is found by accounting for all interaction possibilities of the system of the initial dipole with one extra softer gluon emission depicted in \fig{fig:22tag3} (together with diagrams $B^*$ and $C^*$ not shown in \fig{fig:22tag3}). The corresponding evolution equation is given by 
\begin{eqnarray}\label{evolMA1}
\frac{\partial M(12|12')}{\partial y}=\frac{\bar{\alpha_s}}{2\pi}\int d^2x_3 \left\{
-\frac{1}{2}\left(\frac{x_{13}}{x^2_{13}}-\frac{x_{23}}{x^2_{23}}\right)^2M(12|12')-\frac{1}{2}\left(\frac{x_{13}}{x^2_{13}}-\frac{x_{2'3}}{x^2_{2'3}}\right)^2M(12|12') 
\right.
\end{eqnarray}
\begin{eqnarray}\label{evolMA2}
\left.
+\left(\frac{x_{13}}{x^2_{13}}-\frac{x_{23}}{x^2_{23}}\right)\left(\frac{x_{13}}{x^2_{13}}-\frac{x_{2'3}}{x^2_{2'3}}\right) \left(M(13|13)+M(32|32')+M(13|13)M(32|32')\right. \right.
\end{eqnarray}
\begin{eqnarray}\label{evolMA3}
\left.
- M(32|32') \left\{N(13)+N(13)\right\}
-M(13|13)\left\{N(32)+N(32')\right\}+N(13)N(32')+N(32)N(13) 
\right)
\end{eqnarray}
\begin{eqnarray}\label{evolMA4}
\left.
-\left(\frac{x_{13}}{x^2_{13}}-\frac{x_{23}}{x^2_{23}}\right)\left(0-\frac{x_{2'3}}{x^2_{2'3}}\right)
\left[
M(13|12')\left\{1-N(32)\right\}
+N(32)N(12')\right]\right.
\\
\left.
-\left(\frac{x_{13}}{x^2_{13}}-\frac{x_{23}}{x^2_{23}}\right)\left(\frac{x_{13}}{x^2_{13}}-0\right)
\left[
M(32|12')\left\{1-N(13)\right\}
+N(13)N(12')\right]\right.  \nonumber
\\
\left.
-\left(0-\frac{x_{23}}{x^2_{23}}\right)\left(\frac{x_{13}}{x^2_{13}}-\frac{x_{2'3}}{x^2_{2'3}}\right)
\left[
M(12|13)\left\{1-N(32')\right\}
+N(12)N(32')\right]\right.  \nonumber
\\
\left.
-\left(\frac{x_{13}}{x^2_{13}}-0\right)\left(\frac{x_{13}}{x^2_{13}}-\frac{x_{2'3}}{x^2_{2'3}}\right)
\left[
M(12|32')\left\{1-N(13)\right\}
+N(12)N(13)\right]\right.  \nonumber
\end{eqnarray}
\begin{eqnarray}\label{evolMA5}
\left.
+\left(\frac{x_{13}}{x^2_{13}}-\frac{x_{23}}{x^2_{23}}\right)\left(0-\frac{x_{23}}{x^2_{23}}\right)
\left[
M(13|12')\left\{1-N(32)\right\}
+N(32)N(12')\right]\right.
\\
\left.
+\left(\frac{x_{13}}{x^2_{13}}-\frac{x_{23}}{x^2_{23}}\right)\left(\frac{x_{13}}{x^2_{13}}-0\right)
\left[
M(32|12')\left\{1-N(13)\right\}
+N(13)N(12')\right]\right.  \nonumber
\\
\left.
+\left(0-\frac{x_{2'3}}{x^2_{2'3}}\right)\left(\frac{x_{13}}{x^2_{13}}-\frac{x_{2'3}}{x^2_{2'3}}\right)
\left[
M(12|13)\left\{1-N(32')\right\}
+N(12)N(32')\right]\right.  \nonumber
\\
\left.
+\left(\frac{x_{13}}{x^2_{13}}-0\right)\left(\frac{x_{13}}{x^2_{13}}-\frac{x_{2'3}}{x^2_{2'3}}\right)
\left[
M(12|32')\left\{1-N(13)\right\}
+N(12)N(13)\right]\right.  \nonumber
\end{eqnarray}
\begin{eqnarray}\label{evolMA6}
&&\hspace{0.2cm}\left.
+\left(\frac{x_{23}}{x^2_{23}}\frac{x_{2'3}}{x^2_{2'3}}+\frac{x_{13}}{x^2_{13}}\frac{x_{13}}{x^2_{13}}\right)
M(12|12') 
-\left(\frac{x_{23}}{x^2_{23}}\frac{x_{13}}{x^2_{13}}+\frac{x_{13}}{x^2_{13}}\frac{x_{2'3}}{x^2_{2'3}}\right)
N(12)N(12')
\right. 
\\
&&\left.
-\frac{1}{2}\left(\frac{x_{23}}{x^2_{23}}\frac{x_{23}}{x^2_{23}}+\frac{x_{13}}{x^2_{13}}\frac{x_{13}}{x^2_{13}}\right)
M(12|12') 
+\frac{1}{2}\left(\frac{x_{23}}{x^2_{23}}\frac{x_{13}}{x^2_{13}}+\frac{x_{13}}{x^2_{13}}\frac{x_{23}}{x^2_{23}}\right)
N(12)N(12')
\right. \nonumber
\\
&&\left.
-\frac{1}{2}\left(\frac{x_{2'3}}{x^2_{2'3}}\frac{x_{2'3}}{x^2_{2'3}}+\frac{x_{13}}{x^2_{13}}\frac{x_{13}}{x^2_{13}}\right)
M(12|12') 
+\frac{1}{2}\left(\frac{x_{2'3}}{x^2_{2'3}}\frac{x_{13}}{x^2_{13}}+\frac{x_{13}}{x^2_{13}}\frac{x_{2'3}}{x^2_{2'3}}\right)
N(12)N(12')
\right\}\nonumber
\end{eqnarray}
with $N$ being a solution to the BK equation. 

Each contribution is explained as follows
\begin{itemize}
\item The two terms on the r.h.s of \eq{evolMA1} come from the diagrams $R$ and $R^*$ in \fig{fig:22tag3} and represent the reggeization. Their Kernels are different and correspond to emission and absorption of the softer 
gluon $"3"$  in the amplitude for diagram $R$, and the emission and absorption of the softer 
gluon $"3"$  in the conjugate amplitude for diagram $R^*$.
A factor of $\frac{1}{2}$ appears due to time ordered integral in the light-cone time (for more details see Appendix of Ref.~\cite{Chen:1995pa}).
\item  Terms in \eq{evolMA2} stand for the interaction of each dipole with the target as shown in diagram~$A$. Their Kernel 
corresponds to the emission of the softer gluon $"3"$ in the amplitude and its absorption in the conjugate amplitude. The last non-linear term in \eq{evolMA2} describes the situation where the two dipoles scatter both elastically and inelastically off the target simultaneously. 
\item For all the terms in \eq{evolMA3} we have we same Kernel as for terms in \eq{evolMA2} since emission and 
absorption of the softer gluon is the same. The first and the second terms reflect the situation where one of the dipoles rescatter both 
elastically and inelastically, and the other one rescatters only elastically in either amplitude or conjugate amplitude.
These contributions are not present in $M(13|13)M(32|32')$ since $M$ by the definition includes inelastic part and elastic scattering \emph{both} in the amplitude and the scattering amplitude (in other words, only $N^2$ and not $N$).
 The last term corresponds to the situation where the two dipole rescatter elastically, one in the amplitude and another in conjugate amplitude, and \emph{vice versa}.

\item The same analysis implies in \eq{evolMA4} requiring the proper counting of all interaction patterns in diagrams $B$~(first two lines) and $B^*$~(last two lines) in \fig{fig:22tag3}. 
The zero term in the asymmetric Kernel used here stands for situation where the dipole $"13"("12'")$ is not present by time $\tau=0$ in the conjugate amplitude and thus is omitted as was explained in more details in Section \ref{sec:noevol}.
The last two lines of \eq{evolMA4} correspond to the diagram $B^*$, which is the conjugate of diagram $B$ in 
\fig{fig:22tag3}
\item In \eq{evolMA5} we account for diagram $C$(first two lines) and diagram $C^*$(last two lines) using  the same arguments as for \eq{evolMA4}. The only difference between the corresponding lines of \eq{evolMA4} and \eq{evolMA5}
is the Kernel of the dipole splitting, namely, in \eq{evolMA4} the soft gluon $3$ is present at $\tau=\infty$, and in \eq{evolMA5} is not. If the initial dipole $"12"("12'")$ has the same coordinates in the amplitude and the conjugate amplitude then the corresponding lines of \eq{evolMA4} and \eq{evolMA5} are subject to their mutual cancellation (real-virtual cancellations) and was shown by Chen and Muller \cite{Chen:1995pa}.
 \item  In \eq{evolMA6} we write  the  contributions of diagrams $D$, $E$ and $F$ in \fig{fig:22tag3}. Here one should note that in the crossed dipole splittings ( i.e. $\frac{x_{13}}{x^2_{13}}\frac{x_{23}}{x^2_{23}}$ etc.) the inelastic interaction are suppressed in the large $N_C$ limit, this is the origin of the terms $N(12)N(12')$.
\end{itemize}
The evolution equation should be supplemented by the initial condition $M_0$ given by \eq{MnotA}
\bea\label{initAAA}
M_0(12|12') \,\,&=&\,\,\,1+e^{-x^2_{22'}Q^2_{s}/4}-e^{-x^2_{12}Q^2_{s}/4}-e^{-x^2_{12'}Q^2_{s}/4} 
\eea

One can easily see that due to the equality of the quark coordinate $x_1$ in the amplitude and the conjugate amplitude, some terms cancel each other in the evolution equation  Eqs.~(\ref{evolMA1})-(\ref{evolMA5}). For example, the second line in \eq{evolMA4} cancels  the second line in \eq{evolMA5}. Some further cancellations 
happen because of the optical theorem. The optical theorem states that the total cross section equals twice 
the imaginary part of the scattering amplitude $M(12|12)=2N(12)$. This implies that the last term in \eq{evolMA2} cancels the first term in \eq{evolMA3}. 

The simplified form of the evolution equation Eqs.~(\ref{evolMA1})-(\ref{evolMA5}) is given by 
\begin{eqnarray}\label{evolMA1simple}
\frac{\partial M(12|12')}{\partial y}=\frac{\bar{\alpha_s}}{2\pi}\int d^2x_3 \left\{
-\frac{1}{2}\left(\frac{x_{13}}{x^2_{13}}-\frac{x_{23}}{x^2_{23}}\right)^2M(12|12')-\frac{1}{2}\left(\frac{x_{13}}{x^2_{13}}-\frac{x_{2'3}}{x^2_{2'3}}\right)^2M(12|12') 
\right.
\end{eqnarray}
\begin{eqnarray}\label{evolMA2simple}
\left.
+\left(\frac{x_{13}}{x^2_{13}}-\frac{x_{23}}{x^2_{23}}\right)\left(\frac{x_{13}}{x^2_{13}}-\frac{x_{2'3}}{x^2_{2'3}}\right) \left\{2N(13)+M(32|32')-N(13)N(32')-N(32)N(13)\right\}\right.
\end{eqnarray}
\begin{eqnarray}\label{evolMA3simple}
\left.
-\left(\frac{x_{13}}{x^2_{13}}-\frac{x_{23}}{x^2_{23}}\right)\left(\frac{x_{23}}{x^2_{23}}-\frac{x_{2'3}}{x^2_{2'3}}\right)
\left[
M(13|12')\left\{1-N(32)\right\}
+N(32)N(12')\right]\right.
\\
\left.
-\left(\frac{x_{2'3}}{x^2_{2'3}}-\frac{x_{23}}{x^2_{23}}\right)\left(\frac{x_{13}}{x^2_{13}}-\frac{x_{2'3}}{x^2_{2'3}}\right)
\left[
M(12|13)\left\{1-N(32')\right\}
+N(12)N(32')\right]\right. \nonumber
\end{eqnarray}
\begin{eqnarray}\label{evolMA4simple}
&&\hspace{0.2cm}\left.
-\frac{1}{2}\left(\frac{x_{23}}{x^2_{23}}-\frac{x_{2'3}}{x^2_{2'3}}\right)^2
M(12|12') 
\right\} 
\end{eqnarray}

To see the consistency with the previous studies one can take $2=2'$ and use the optical theorem to 
recover the BK equation. In this case  \eq{evolMA3simple} and \eq{evolMA4simple} vanish, Kernels in \eq{evolMA1simple} and \eq{evolMA2simple} become equal and we end up with 
with the evolution equation  
\begin{eqnarray}\label{BK2}
\frac{\partial( 2 N(12))}{\partial y}=\frac{\bar{\alpha_s}}{2\pi}\int d^2x_3 \left(\frac{x_{13}}{x^2_{13}}-\frac{x_{23}}{x^2_{23}}\right)^2
\left\{2N(13)+2N(32)-2N(13)N(32)\right\}
\end{eqnarray}
with the initial condition directly obtained from ~(see \eq{MnotA})
\begin{eqnarray}\label{BK2init}
 M_{0}(12|12')=1+e^{-x^2_{22'}Q^2_{so}/4}-e^{-x^2_{12}Q^2_{so}/4}-e^{-x^2_{12'}Q^2_{so}/4}
\end{eqnarray}
 for $2=2'$ as follows 
$M_{0}(12|12)=2\left(1-e^{-x^2_{12}Q^2_{so}/4}\right)=2N_0(12)$.
Thus the evolution equation is the BK equation with the correct initial condition.

Next step is to study  general properties and to find a solution to  the evolution equation  Eqs.~(\ref{evolMA1simple})-(\ref{evolMA4simple}). At this point we make a digression and go back to the classical expression for the single inclusive production cross section Eqs.~(\ref{line1})-(\ref{line5}).
 As one can easily see,  the cancellation of the emissions with the "spectator" quark~(antiquark) found  by Kovchegov \cite{Kovchegov:2001ni} in the form of Eqs.~(\ref{Classline1})-(\ref{Classline4}) can happen only if 
$M_0(12|10)=N_0(12)+N_0(10)-N_0(20)$, where in the last term $20=(10)-(12)$.
The initial condition \eq{BK2init} indeed satisfies this condition. If one wants to include the evolution effects, the function $M_0$ in  Eqs.~(\ref{line1})-(\ref{line5}) should be replaced by the solution to the evolution equation 
Eqs.~(\ref{evolMA1simple})-(\ref{evolMA4simple}) with the initial condition given by \eq{BK2init}.
This suggests that $M(12|12')$ should also have this property to result into the cancellation of the emissions and the interactions with the "spectator" quark~(antiquark) line.
The straightforward substitution of 
\begin{eqnarray}\label{M1212}
M(12|12')=N(12)+N(12')-N(22')
\end{eqnarray}
into Eqs.~(\ref{evolMA1simple})-(\ref{evolMA4simple}) reveals
\small
\begin{eqnarray}\label{evolMA1sub}
\frac{\partial (N(12)+N(12')-N(22'))}{\partial y}=\frac{\bar{\alpha_s}}{2\pi}\int d^2x_3 \left\{
-\frac{1}{2}
\left(
 \left(\frac{x_{13}}{x^2_{13}}-\frac{x_{23}}{x^2_{23}}\right)^2+\left(\frac{x_{13}}{x^2_{13}}-\frac{x_{2'3}}{x^2_{2'3}}\right)^2 
\right) (N(12)+N(12')-N(22')) 
\right.
\end{eqnarray}
\begin{eqnarray}\label{evolMA2sub}
\left.
+\left(\frac{x_{13}}{x^2_{13}}-\frac{x_{23}}{x^2_{23}}\right)\left(\frac{x_{13}}{x^2_{13}}-\frac{x_{2'3}}{x^2_{2'3}}\right) \left\{2N(13)+(N(32)+N(32')-N(22'))-N(13)N(32')-N(32)N(13)\right\}\right.
\end{eqnarray}
\begin{eqnarray}\label{evolMA3sub}
\left.
-\left(\frac{x_{13}}{x^2_{13}}-\frac{x_{23}}{x^2_{23}}\right)\left(\frac{x_{23}}{x^2_{23}}-\frac{x_{2'3}}{x^2_{2'3}}\right)
\left[
\left( N(13)+N(12')-N(32')\right)
\left\{1-N(32)\right\}
+N(32)N(12')\right]\right.
\\
\left.
-\left(\frac{x_{2'3}}{x^2_{2'3}}-\frac{x_{23}}{x^2_{23}}\right)\left(\frac{x_{13}}{x^2_{13}}-\frac{x_{2'3}}{x^2_{2'3}}\right)
\left[
\left(N(12)+N(13)-N(32)\right)\left\{1-N(32')\right\}
+N(12)N(32')\right]\right. \nonumber
\end{eqnarray}
\begin{eqnarray}\label{evolMA4sub}
&&\hspace{0.2cm}\left.
-\frac{1}{2}\left(\frac{x_{23}}{x^2_{23}}-\frac{x_{2'3}}{x^2_{2'3}}\right)^2
\left(N(12)+N(12')-N(22')\right) 
\right\} 
\end{eqnarray}

\normalsize
which is just a linear combination of three BK equations for initial dipoles with coordinates $12$, $12'$ and $22'$

\begin{eqnarray}\label{evolMA1comb}
\frac{\partial (N(12)+N(12')-N(22'))}{\partial y}=\frac{\bar{\alpha_s}}{2\pi}\int d^2x_3 \left\{
 \left(\frac{x_{13}}{x^2_{13}}-\frac{x_{23}}{x^2_{23}}\right)^2  \left\{N(13)+N(32)-N(12)-N(13)N(32)\right\}
\right.
\end{eqnarray}
\begin{eqnarray}\label{evolMA2comb}
\left.
+\left(\frac{x_{13}}{x^2_{13}}-\frac{x_{2'3}}{x^2_{2'3}}\right)^2  \left\{N(13)+N(32')-N(12')-N(13)N(32')\right\}
\right.
\end{eqnarray}
\begin{eqnarray}\label{evolMA3comb}
\left.
-\left(\frac{x_{23}}{x^2_{23}}-\frac{x_{2'3}}{x^2_{2'3}}\right)^2  \left\{N(32)+N(32')-N(22')-N(32)N(32')\right\}
\right\} 
\end{eqnarray}
This way we show explicitly that indeed the cancellations claimed by Kovchegov and Tuchin \cite{Kovchegov:2001sc}
hold to any order of  evolution, allowing for the simple description of the single inclusive cross section
in terms of the combination $2N(x)-N^2(x)$, which can be interpreted as the scattering amplitude $N_G(x)$ of an adjoint("gluonic") dipole of size~$x$.

In the next Section we examine the Abramovsky-Gribov-Kancheli cutting rules in pQCD using the formalism we developed for the calculation of the single production cross section.

\section{Abramovsky-Gribov-Kancheli cutting rules}\label{sec:agk}
More than 30 years ago Abramovsky,Gribov and Kancheli \cite{Abramovsky:1973fm} analyzed the relative contributions
of processes with different multiplicity of the produced particles to the total cross section of the multiple Pomeron exchange in the framework of the 
Reggeon Field Theory. They found that the contributions of  different unitarity cuts, corresponding to different particle multiplicities are proportional to each other (AGK cutting rules), and their relative coefficients lead to some peculiar cancellations, called AGK cancellations for the single jet production. The most interesting case is the two Pomeron exchange, where the AGK cancellation can be explained as follows. The total cross section is given by the sum of all possible  unitarity cuts. The Pomeron is non-local object, so it can be also cut enlarging the number of  cut possibilities. This way one obtains two distinct objects - cut and uncut Pomerons, which are to be treated separately as far as multiplicity is concerned. 
 The simplest example for this is the two Pomeron exchange depicted in \fig{fig:agk}.
\FIGURE[h]{ \centerline{
 \epsfig{file=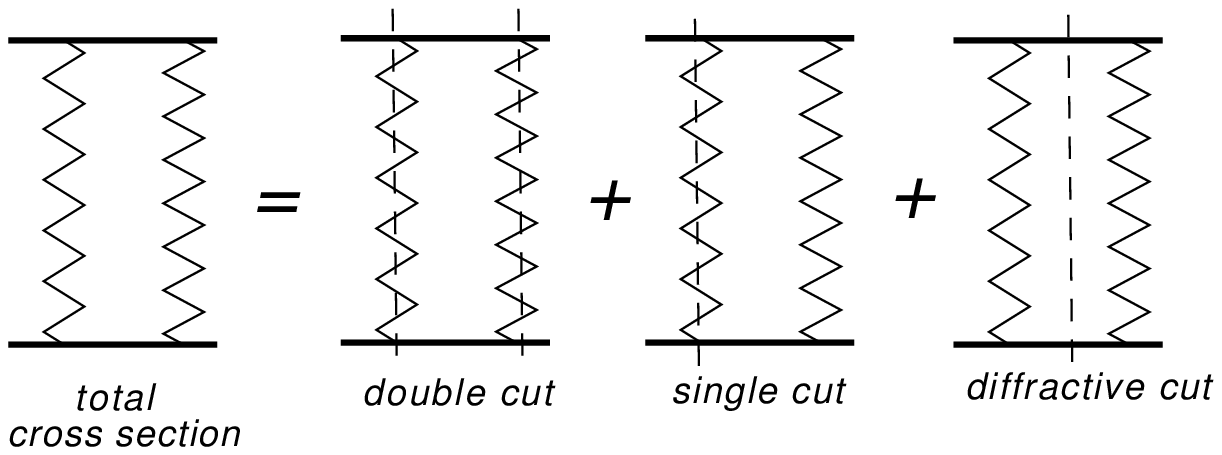,width=110mm,height=35mm} } \caption{
 The total cross section of two Pomeron exchange in terms of contributions from different $s$-channel unitarity cuts. The double cut diagram describes the process with the multiplicity  two times larger   the multiplicity for one Pomeron exchange (see \protect\fig{cutpom}), the single cut diagrams leads to a process with the same multiplicity as one Pomeron exchange (see \protect\fig{cutpom}) and the diffractive cut diagram gives the expression for a diffractive production with small multiplicity. }\label{fig:agk}}
As one can see we have here three different processes: $a)$ \emph{double cut}, where the two Pomerons are cut.
 This brings two times multiplicity of one cut Pomeron; $b)$ \emph{single cut}, one Pomeron appears cut and the other one is uncut. The multiplicity of this process equals  the multiplicity of one cut Pomeron; $c)$ \emph{diffractive cut}, here two Pomerons are uncut and no particles are produced. The multiplicity of this process is very small. The relative weights of three processes are calculated as follows. There only one possibility to cut two Pomerons at the same time, each cut Pomeron is twice the uncut Pomeron exchange
( due to the unitarity condition $\displaystyle{ \not \hspace{-0.1cm}P}=2P$ ), thus we obtain $4P^2$. 
For the single cut we have two possibilities: to cut one of two Pomerons, this must be multiplied by the a factor of $2$ for putting the uncut Pomeron either in the amplitude or the conjugate amplitude. The  resulting combinatorial 
coefficient of $4$ is to be multiplied by a factor $2$ for the normalization of the cut Pomeron 
($\displaystyle{ \not \hspace{-0.1cm}P}=2P$) as well as a minus sign originating from the fact that uncut Pomeron is the imaginary part of the scattering amplitude. Summarizing all this, one gets $-8P^2$  from the single cut.
Finally,  in the diffractive cut the combinatorial coefficient is $2$ for interchanging the two Pomerons in the amplitude and the conjugate amplitude. All three cuts add up to $4P^2-8P^2+2P^2=-2P^2=\sigma^{(2)}_{tot}$, as expected.

This set of the AGK coefficients results into cancellation of the single jet production as shown in 
\fig{fig:agk-cancel}

\FIGURE[h]{ \centerline{
 \epsfig{file=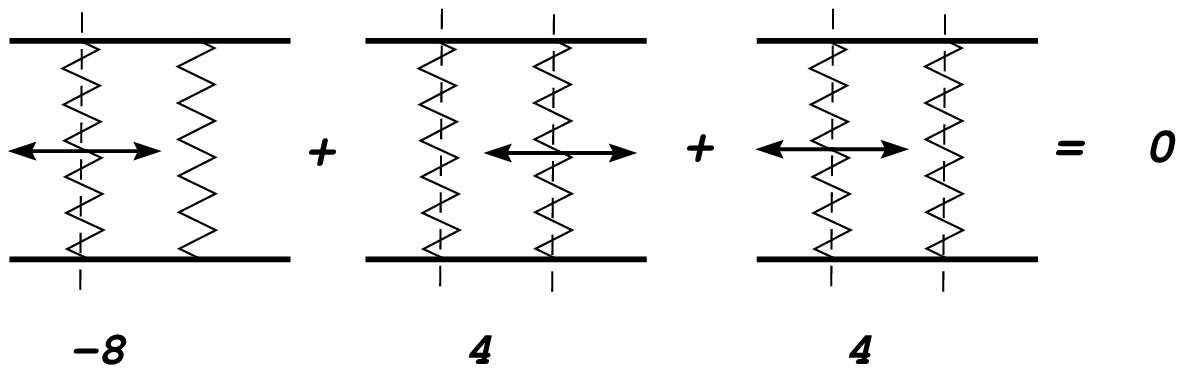,width=110mm} } \caption{
 The cancellation of the single jet production for two Pomeron exchange.}\label{fig:agk-cancel}} 
The jet can be produced from either of two cut Pomerons in the double cut, and only from the cut Pomeron in the single cut, resulting into the cancellation of the single jet production in the two Pomeron exchange.
In this Section we treat the contributions from the different unitarity cuts in pQCD in the framework of the color dipole model.
The goal of this study is to see how the AGK cancellations happen in the color dipole model and check in this way the validity of the AGK cutting rules in pQCD.

In the original derivation of the AGK cutting rules \cite{Abramovsky:1973fm} some basic assumptions were made, namely,
\begin{itemize}
\item Pomeron is a proper asymptotics in the high energy scattering;
\item there are no "half-cut" Pomerons, the contributions where a  Pomeron is not cut entirely are suppressed exponentially in the invariant mass;
\item there is no contribution to the multiplicity from the cut of the vertices;
\item all the vertices for various cuts are the same and real.
\end{itemize}
In the AGK paper \cite{Abramovsky:1973fm} all these  assumptions have been proved in the superconverged
field theories, since they guarantee  that all scattering amplitudes decrease in the region of large masses.
As has been shown in Ref.~\cite{GLR} in the leading $log(1/x)$ approximation of perturbative QCD, the amplitudes fall in the region of high mass fast enough to hope that AGK cutting rules are correct. However, in addition to the assumptions that discussed above, it was also assumed in Ref.~\cite{Abramovsky:1973fm} that the multiparticle production is described by the same set of diagram as the total cross section. This assumption
was doubtful even in a superconverged theory, since it was found that there is an example of the diagrams that does not contribute to the total cross section, but could lead to the multiparticle production ( the AFS-type of diagrams \cite{AFS} ).  Therefore, our main goal is to find what types of diagrams indeed contribute to the total and inelastic cross sections and to calculate the relation between different production processes.

Our strategy is as follows. We use the formalism developed in the previous Sections for deriving cutting rules
in pQCD, then we analyze the single inclusive production of the two Pomeron exchange and check how the emission of the gluon changes the cutting rules. The process we discuss is  the dipole scattering on the target via two Pomeron exchange: $a)$  two Pomerons couple to the dipole; $b)$ one Pomeron couples to the dipole and then splits into two Pomerons. The first case trivially satisfies AGK cutting rules and thus is of no interest. We focus on the second case depicted in \fig{fig:process}. At this stage we do not consider any gluon production and  want to analyze
only the
 two Pomeron exchange in terms of the contributions of different cuts.

\FIGURE[h]{ \centerline{
 \epsfig{file=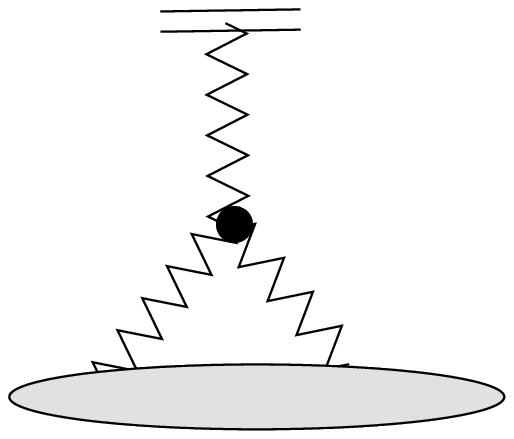,width=50mm} } \caption{
 We check the validity the AGK cutting rules for the process of one to two Pomeron splitting.}
\label{fig:process}}

Schematically, this process can be written as 
\begin{eqnarray}\label{2Pschem}
\sigma=\int P(Y-Y_0)\otimes \Gamma_{3P} \otimes P^2(Y_0-0)
\end{eqnarray}
where $P(Y-Y_0)$ is the Pomeron amplitude and  $\Gamma_{3P} $ is the triple Pomeron vertex. In the case of the virtual
photon scattering on the target this is given by 

\begin{eqnarray}\label{process}
\sigma=\int d^2x_{\tilde{1}}d^2x_{\tilde{0}}d\tilde{z} \;d^2x_{1}d^2x_{0} \;|\psi^{\gamma^*\rightarrow q\bar{q}}(x_{\tilde{1}\tilde{0}},\tilde{z})|^2   n(x_{\tilde{1}},x_{\tilde{0}}|x_{1},x_{0};Y-Y_0) \sigma_{2P}^{q\bar{q}}(x_{1},x_0)
\end{eqnarray}
 where $\psi^{\gamma^*\rightarrow q\bar{q}}(x_{\tilde{1}\tilde{0}},\tilde{z})$ is the wave function of the virtual photon being split into $q\bar{q}$ pair, $n(x_{\tilde{1}},x_{\tilde{0}}|x_{1},x_{0};Y-Y_0)$ is the dipole density, which is the solution to the linear evolution equation. The triple Pomeron vertex and the interaction with the target is encoded into $\sigma_{2P}^{q\bar{q}}(x_{1},x_0)$.  Using the formalism developed in the previous 
Sections we can readily write an expression for $\sigma_{2P}^{q\bar{q}}(x_{1},x_0)$ as 

\begin{eqnarray}\label{sigma2P}
\sigma_{2P}^{q\bar{q}}(x_{1},x_0)=\frac{\bar{\alpha}_s}{2\pi}\int^{Y_0}_0 dy \int d^2 x_2 
\left(\frac{x_{12}}{x^2_{12}}-\frac{x_{02}}{x^2_{02}}\right)^2
\left\{M_{lin}(12|12)M_{lin}(20|20)+\frac{1}{2}\left(M_{lin}(12|12)\right)^2 
\right.  \nonumber
\\
\left.
+\frac{1}{2}\left(M_{lin}(20|20)\right)^2
-\frac{1}{2}\left(M_{lin}(10|10)\right)^2-\{M_{lin}(12|12)+M_{lin}(20|20)\}\{2N_{\mathtt{BFKL}}(12)+2N_{\mathtt{BFKL}}(20)\}
\right. 
\\
\left.
+M_{lin}(10|10)2N_{\mathtt{BFKL}}(10)
+2N_{\mathtt{BFKL}}(12)N_{\mathtt{BFKL}}(20) +\left(N_{\mathtt{BFKL}}(12)\right)^2
+\left(N_{\mathtt{BFKL}}(20)\right)^2
-\left(N_{\mathtt{BFKL}}(10)\right)^2
\right\} \nonumber
\end{eqnarray}
where $N_{\mathtt{BFKL}}$ and $M_{lin}$ are solutions to the BFKL equation and some linear evolution equation 
as explained below.

In \eq{sigma2P} the triple Pomeron vertex is represented by the initial dipole $"10"$ which splits into two new 
dipoles $"12"$ and $"02"$, each of them can interact with the target elastically by the BFKL amplitude
$N_{BFKL}$  and inelastically by the inelastic cross section $M_{lin}$.
If the same dipole interacts twice inelastically we put a factor of $\frac{1}{2}$ for the
identity of the interactions. 
We have to take into account all possible splittings of the initial dipole as shown in \fig{fig:22tag3}, since  most of the splittings are canceled by the real-virtual cancellations and we are left with diagrams $A$, $R$ and $R^*$.
The inelastic cross section $M_{lin}$  is the solution to the linear evolution equation, obtained from the non-linear evolution equation Eqs.~(\ref{evolMA1})-(\ref{evolMA6}) retaining only the its linear part. In the contrary to the 
the non-linear case we do not include multiple rescattering in the initial condition for $M_{lin}$, which now consists of only two $t$-channel gluon exchange. In this case 
the total cross section equals the inelastic cross section and $M_{lin}$ is associated with a cut Pomeron. For the non-linear case, where the multiple rescattering is taken into account  the solution to Eqs.~(\ref{evolMA1})-(\ref{evolMA6}) includes both cut and uncut Pomeron exchanges, which can be extracted by expanding the total cross section in terms of the Pomerons as we show below.
As we have already mentioned the linear evolution equation for $M_{lin}$ is readily obtained from  Eqs.~(\ref{evolMA1})-(\ref{evolMA6}) neglecting the non-linear terms as   
\begin{eqnarray}\label{evolMlin1}
\frac{\partial M_{lin}(12|12')}{\partial y}=\frac{\bar{\alpha_s}}{2\pi}\int d^2x_3 \left\{
-\frac{1}{2}\left(\frac{x_{13}}{x^2_{13}}-\frac{x_{23}}{x^2_{23}}\right)^2M_{lin}(12|12')-\frac{1}{2}\left(\frac{x_{13}}{x^2_{13}}-\frac{x_{2'3}}{x^2_{2'3}}\right)^2M_{lin}(12|12') 
\right.
\end{eqnarray}
\begin{eqnarray}\label{evolMlin2}
\left.
+\left(\frac{x_{13}}{x^2_{13}}-\frac{x_{23}}{x^2_{23}}\right)\left(\frac{x_{13}}{x^2_{13}}-\frac{x_{2'3}}{x^2_{2'3}}\right) \left\{M_{lin}(13|13)+M_{lin}(32|32')\right\} \right.
\end{eqnarray}
\begin{eqnarray}\label{evolMlin4}
\left.
-\left(\frac{x_{13}}{x^2_{13}}-\frac{x_{23}}{x^2_{23}}\right)\left(0-\frac{x_{2'3}}{x^2_{2'3}}\right)
M_{lin}(13|12')\right.
\\
\left.
-\left(\frac{x_{13}}{x^2_{13}}-\frac{x_{23}}{x^2_{23}}\right)\left(\frac{x_{13}}{x^2_{13}}-0\right)
M_{lin}(32|12')\right.  \nonumber
\\
\left.
-\left(0-\frac{x_{23}}{x^2_{23}}\right)\left(\frac{x_{13}}{x^2_{13}}-\frac{x_{2'3}}{x^2_{2'3}}\right)
M_{lin}(12|13)\right.  \nonumber
\\
\left.
-\left(\frac{x_{13}}{x^2_{13}}-0\right)\left(\frac{x_{13}}{x^2_{13}}-\frac{x_{2'3}}{x^2_{2'3}}\right)
M_{lin}(12|32')\right.  \nonumber
\end{eqnarray}
\begin{eqnarray}\label{evolMlin5}
\left.
+\left(\frac{x_{13}}{x^2_{13}}-\frac{x_{23}}{x^2_{23}}\right)\left(0-\frac{x_{23}}{x^2_{23}}\right)
M_{lin}(13|12')\right.
\\
\left.
+\left(\frac{x_{13}}{x^2_{13}}-\frac{x_{23}}{x^2_{23}}\right)\left(\frac{x_{13}}{x^2_{13}}-0\right)
M_{lin}(32|12')\right.  \nonumber
\\
\left.
+\left(0-\frac{x_{2'3}}{x^2_{2'3}}\right)\left(\frac{x_{13}}{x^2_{13}}-\frac{x_{2'3}}{x^2_{2'3}}\right)
M_{lin}(12|13)\right.  \nonumber
\\
\left.
+\left(\frac{x_{13}}{x^2_{13}}-0\right)\left(\frac{x_{13}}{x^2_{13}}-\frac{x_{2'3}}{x^2_{2'3}}\right)
M_{lin}(12|32')\right.  \nonumber
\end{eqnarray}
\begin{eqnarray}\label{evolMlin6}
&&\hspace{0.2cm}\left.
+\left(\frac{x_{23}}{x^2_{23}}\frac{x_{2'3}}{x^2_{2'3}}+\frac{x_{13}}{x^2_{13}}\frac{x_{13}}{x^2_{13}}\right)
M_{lin}(12|12') 
\right. 
\\
&&\left.
-\frac{1}{2}\left(\frac{x_{23}}{x^2_{23}}\frac{x_{23}}{x^2_{23}}+\frac{x_{13}}{x^2_{13}}\frac{x_{13}}{x^2_{13}}\right)
M_{lin}(12|12') 
\right. \nonumber
\\
&&\left.
-\frac{1}{2}\left(\frac{x_{2'3}}{x^2_{2'3}}\frac{x_{2'3}}{x^2_{2'3}}+\frac{x_{13}}{x^2_{13}}\frac{x_{13}}{x^2_{13}}\right)
M_{lin}(12|12') \right\}\nonumber
\end{eqnarray}
with the initial condition given by 
\begin{eqnarray}\label{initLin}
M_{lin_0}(12|12')=\frac{\bar{\alpha}_s}{8\pi}\int d^2k \frac{ (e^{-ikx_1}-e^{-ikx_2})(e^{ikx_1}-e^{ikx_{2'}})}{k^4}
IF(k^2)
\end{eqnarray}
where $IF(k^2)$ is the impact factor of the target.
 
The terms on r.h.s of \eq{evolMlin1} are the contributions from the diagrams $R$ and $R^*$ in \fig{fig:22tag3}, \eq{evolMlin2} comes from diagram $A$, \eq{evolMlin4} account for diagrams $B$ and $B^*$, \eq{evolMlin5} stands  
for diagrams $C$ and $C^*$, and, finally,  \eq{evolMlin4} describes diagrams $D$, $E$ and $F$.  A more detailed explanation is presented in Section~\ref{sec:evol}.

Eqs.~(\ref{evolMlin1})-(\ref{evolMlin6}) can be written in a compact form   as
\begin{eqnarray}\label{evolMlin1comp}
\frac{\partial M_{lin}(12|12')}{\partial y}=\frac{\bar{\alpha_s}}{2\pi}\int d^2x_3 \left\{
-\frac{1}{2}\left(\frac{x_{13}}{x^2_{13}}-\frac{x_{23}}{x^2_{23}}\right)^2M_{lin}(12|12')-\frac{1}{2}\left(\frac{x_{13}}{x^2_{13}}-\frac{x_{2'3}}{x^2_{2'3}}\right)^2M_{lin}(12|12') 
\right.
\end{eqnarray}
\begin{eqnarray}\label{evolMlin2comp}
\left.
+\left(\frac{x_{13}}{x^2_{13}}-\frac{x_{23}}{x^2_{23}}\right)\left(\frac{x_{13}}{x^2_{13}}-\frac{x_{2'3}}{x^2_{2'3}}\right) \left\{M_{lin}(13|13)+M_{lin}(32|32')\right\} \right.
\end{eqnarray}
\begin{eqnarray}\label{evolMlin3comp}
\left.
-\left(\frac{x_{13}}{x^2_{13}}-\frac{x_{23}}{x^2_{23}}\right)\left(\frac{x_{23}}{x^2_{23}}-\frac{x_{2'3}}{x^2_{2'3}}\right)
M_{lin}(13|12')
-\left(\frac{x_{2'3}}{x^2_{2'3}}-\frac{x_{23}}{x^2_{23}}\right)\left(\frac{x_{13}}{x^2_{13}}-\frac{x_{2'3}}{x^2_{2'3}}\right)
M_{lin}(12|13)\right.  \nonumber
\end{eqnarray}
\begin{eqnarray}\label{evolMlin4comp}
&&\hspace{0.2cm}\left.
-\frac{1}{2}\left(\frac{x_{23}}{x^2_{23}}-\frac{x_{2'3}}{x^2_{2'3}}\right)^2
M_{lin}(12|12') 
 \right\}
\end{eqnarray}

For $x_2=x_{2'}$ all terms in Eqs.~(\ref{evolMlin3comp})-(\ref{evolMlin4comp}) cancel out and the linear evolution 
equation reduces to 
\begin{eqnarray}\label{evolM1BFKL}
\frac{\partial M_{lin}(12|12)}{\partial y}=\frac{\bar{\alpha_s}}{2\pi}\int d^2x_3 
\left(\frac{x_{13}}{x^2_{13}}-\frac{x_{23}}{x^2_{23}}\right)^2\left\{-\frac{2}{2}M_{lin}(12|12)+M_{lin}(13|13)+M_{lin}(32|32) \right\}
\end{eqnarray}
which is BFKL equation because of the unitarity condition $M_{lin}(12|12)=2N_{lin}(12)$.
As in the case of the non-linear equation it is easy to see from Eqs.~(\ref{evolMlin1comp})-(\ref{evolMlin4comp}) that its solution $M_{lin}(12|12')$ can be expressed through solution to BFKL equation as 
\begin{eqnarray}\label{unitGen}
M_{lin}(12|12')=N_{BFKL}(12)+N_{BFKL}(12')-N_{BFKL}(22')
\end{eqnarray}
This is a generalized form of the optical theorem, which relates the total cross section $M_{lin}(12|12')$
of scattering of the dipole $"12"("12'")$ to the elastic scattering amplitude of the dipoles $"12"$, $"12'"$ and $"12-12'"$.

We are interested in the contributions to the total cross section, which have different multiplicities as shown in 
\fig{fig:agkqcd}
\FIGURE[h]{ \centerline{
 \epsfig{file=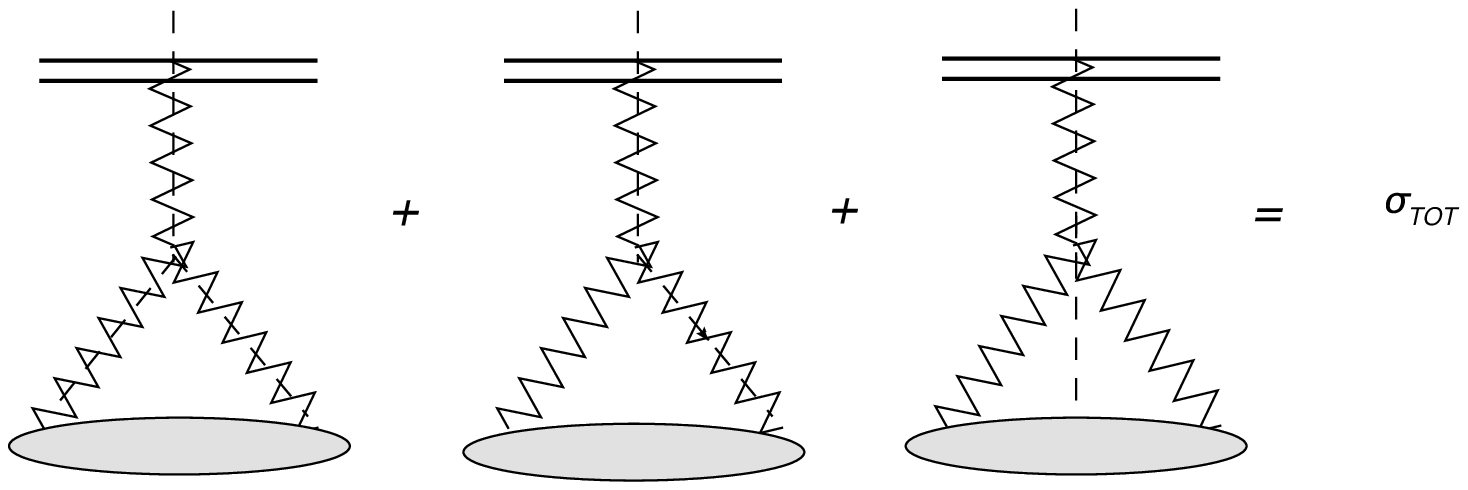,width=110mm} } \caption{
 The contributions to the total cross section having different multiplicities.}
\label{fig:agkqcd}}

We go back to the expression for the triple Pomeron vertex with interaction \eq{sigma2P}
and read out the contributions of the different cuts as follows.
\begin{center}
\begin{eqnarray}
\begin{tabular}{cc}\label{cuts}
Double Cut &
 $\frac{1}{2}\left(M_{lin}(12|12)+M_{lin}(20|20)\right)^2-\frac{1}{2}M_{lin}(10|10)$ 
\vspace{0.2cm}
\\
Single Cut &
 $-\left(M_{lin}(12|12)+M_{lin}(20|20)\right)
\left(2N_{\mathtt{BFKL}}(12)+2N_{\mathtt{BFKL}}(20)\right)
+M_{lin}(10|10)2N_{\mathtt{BFKL}}(10)$ 
\vspace{0.2cm}
\\
Diffractive Cut &
 $\frac{2}{2}\left(N_{\mathtt{BFKL}}(12)+N_{\mathtt{BFKL}}(02)\right)^2-\frac{2}{2}N^2_{\mathtt{BFKL}}(10)$ 
\end{tabular}
\end{eqnarray}
\end{center}
One can easily see using the unitarity relation \eq{unitGen}  
$M_{lin}(ij|ij)=2N_{\mathtt{BFKL}}(ij)$ both the double and  the single cut contributions are proportional
to the diffractive term  as expected from the AGK cutting rules.
\begin{center}
\begin{eqnarray}
\begin{tabular}{cc}\label{cutsDone}
Double Cut &
 $2\left(N_{\mathtt{BFKL}}(12)+N_{\mathtt{BFKL}}(02)\right)^2-2N^2_{\mathtt{BFKL}}(10)$ 
\vspace{0.2cm}
\\
Single Cut &
 $-4\left(N_{\mathtt{BFKL}}(12)+N_{\mathtt{BFKL}}(02)\right)^2+4N^2_{\mathtt{BFKL}}(10)$ 
\vspace{0.2cm}
\\
Diffractive Cut &
 $\left(N_{\mathtt{BFKL}}(12)+N_{\mathtt{BFKL}}(02)\right)^2-N^2_{\mathtt{BFKL}}(10)$ 
\end{tabular}
\end{eqnarray}
\end{center}
It is instructive to demonstrate how terms in \eq{cutsDone} add up to the total cross section. The easiest way to see this is to write down the Glauber expression for the total cross section in \eq{sigma2P} as follows
\begin{eqnarray}\label{sigma2PGlaub}
\sigma_{2P}^{q\bar{q}}(x_{1},x_0)=\frac{\bar{\alpha}_s}{2\pi}\int^{Y_0}_0 dy \int d^2 x_2 
\left(\frac{x_{12}}{x^2_{12}}-\frac{x_{02}}{x^2_{02}}\right)^2
\left\{
2\left(1-e^{\frac{1}{2}(\sigma^{BA}(12)+\sigma^{BA}(20))}
\right)
-
2\left(1-e^{\frac{1}{2}\sigma^{BA}(10)}
\right)\right\} \nonumber
\end{eqnarray}
where we used the notation of Appendix~\ref{sec:B} and absorbed for simplicity $T(b;R_A)$ in the definition 
of $\sigma^{BA}$. The first term in the brackets in \eq{sigma2PGlaub} is the total cross section of the scattering of $q\bar{q}g$ system after the soft gluon emission~(dipole splitting) happened, this term comes from diagram $A$ if \fig{fig:real-virt}. The second terms in \eq{sigma2PGlaub} corresponds to the situation where the gluon is both emitted and absorbed before the scattering takes place~(see diagrams $R$ and $R^*$ in \fig{fig:real-virt}).
It is easy to see that the expansion of the expression in the brackets in \eq{sigma2PGlaub} to the second order in  $\sigma^{BA}$ with the help of 
$\sigma^{BA}=2N^{BA}$ gives the sum of terms in \eq{cutsDone}.  

The coefficients in \eq{cutsDone} show that we fully reproduce the AGK cutting in this case. However, the situation changes 
when one emits a gluon from the triple Pomeron vertex. In this case the real-virtual cancellations  does not happen and one should expect non-trivial contributions from different cuts as  discussed in the next Section.

\section{AGK rules for a production from vertex}\label{sec:agkinc}
In deriving  the triple Pomeron vertex  \eq{sigma2P} we used the real-virtual cancellations, where the real gluon emissions were cancelled by the virtual emissions. This does not happen  for the real gluon production 
(see  \fig{fig:process1}) and  we have to go back to extract only the real emission terms.  
\FIGURE[h]{ \centerline{
 \epsfig{file=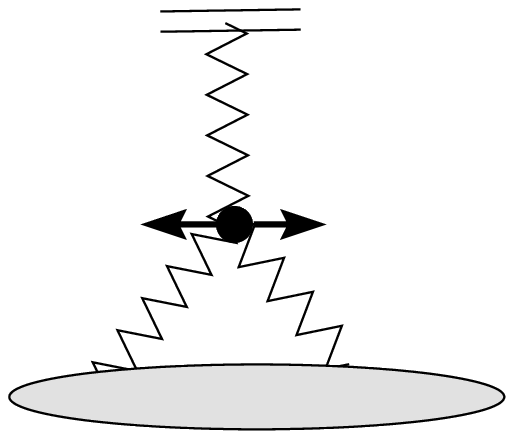,width=50mm} } \caption{
 The emission from the vertex in one to two Pomerons splitting.}
\label{fig:process1}}

As in the case of no gluon production, this process can be written as 
\begin{eqnarray}\label{2Pschem1}
\frac{d\sigma}{dY_0\;d^2k}=\int P(Y-Y_0)\otimes \Gamma_{3P}(k,Y_0) \otimes P^2(Y_0-0)
\end{eqnarray}
where $P(Y-Y_0)$ is the Pomeron amplitude and  $\Gamma_{3P} $ is the triple Pomeron vertex with one gluon emission. In the case of the virtual
photon scattering on the target this is given by 

\begin{eqnarray}\label{process1}
\frac{d\sigma}{dY_0\;d^2k}=\int d^2x_{\tilde{1}}d^2x_{\tilde{0}}d\tilde{z} \;d^2x_{1}d^2x_{0} \;|\psi^{\gamma^*\rightarrow q\bar{q}}(x_{\tilde{1}\tilde{0}},\tilde{z})|^2   n_{lin}(x_{\tilde{1}},x_{\tilde{0}}|x_{1},x_{0};Y-Y_0) \frac{d\sigma_{2P}^{q\bar{q}}(x_{1},x_0)}{dY_0 d^2k}
\end{eqnarray}
 where $\psi^{\gamma^*\rightarrow q\bar{q}}(x_{\tilde{1}\tilde{0}},\tilde{z})$ is the wave function of the virtual photon being split into $q\bar{q}$ pair, $n_{lin}(x_{\tilde{1}},x_{\tilde{0}}|x_{1},x_{0},Y-Y_0)$ is the dipole density, which is the solution to the linear evolution equation. 

All the information about the contributions of the different unitarity cuts is encoded in the differential cross section  $\frac{d\sigma_{2P}^{q\bar{q}}(x_{1},x_0)}{dY_0 d^2k}$.  We can readily write it following the  strategy we used before in deriving the single inclusive cross section Eqs.~(\ref{line1})-(\ref{line5}) in Section~\ref{sec:noevol}. 

\begin{eqnarray}\label{agkV1}
 \frac{d\sigma_{2P}^{q\bar{q}}(x_{1},x_0)}{dY_0 d^2k}=\frac{\bar{\alpha}_s}{2\pi}\frac{1}{(2\pi)^2}
\int d^2 x_2 \; d^2 x_{2'}e^{-ik(x_2-x_{2'})}\left(\left(\frac{x_{12}}{x^2_{12}}-\frac{x_{02}}{x^2_{02}}\right)
\left(\frac{x_{12'}}{x^2_{12'}}-\frac{x_{02'}}{x^2_{02'}}\right)
 \{
\frac{1}{2}M^2_{lin}(12|12')+\frac{1}{2}M^2_{lin}(20|2'0) \right.  \hspace{0.7cm}
\\
 \left.   
+M_{lin}(12|12')M_{lin}(20|2'0)
-(M_{lin}(12|12')+M_{lin}(20|2'0))
(N_{\mathtt{BFKL}}(12)+N_{\mathtt{BFKL}}(12')+N_{\mathtt{BFKL}}(20)+N_{\mathtt{BFKL}}(2'0))
\right. \nonumber
\\
  \left. 
+N_{\mathtt{BFKL}}(12)N_{\mathtt{BFKL}}(2'0)+N_{\mathtt{BFKL}}(20)N_{\mathtt{BFKL}}(12')
+N_{\mathtt{BFKL}}(12)N_{\mathtt{BFKL}}(12')
+N_{\mathtt{BFKL}}(20)N_{\mathtt{BFKL}}(2'0)
 \}\right. \nonumber
\end{eqnarray}
\begin{eqnarray}\label{agkV2}
  -\left(\frac{x_{12}}{x^2_{12}}-\frac{x_{02}}{x^2_{02}}\right)
\left(0-\frac{x_{02'}}{x^2_{02'}}\right)
\{
-M_{lin}(12|10)\left(N_{\mathtt{BFKL}}(20)+N_{\mathtt{BFKL}}(12)+N_{\mathtt{BFKL}}(10)\right)
 \nonumber\\
 +\frac{1}{2}M^2_{lin}(12|10)+N_{\mathtt{BFKL}}(12)N_{\mathtt{BFKL}}(10)+N_{\mathtt{BFKL}}(20)N_{\mathtt{BFKL}}(10)
 \} \nonumber
\\
-\left(\frac{x_{12}}{x^2_{12}}-\frac{x_{02}}{x^2_{02}}\right)
\left(\frac{x_{12'}}{x^2_{12'}}-0\right)
\{
-M_{lin}(20|10)\left(N_{\mathtt{BFKL}}(12)+N_{\mathtt{BFKL}}(20)+N_{\mathtt{BFKL}}(10)\right)
\nonumber
 \\
 +\frac{1}{2}M^2_{lin}(20|10)+N_{\mathtt{BFKL}}(12)N_{\mathtt{BFKL}}(10)+N_{\mathtt{BFKL}}(20)N_{\mathtt{BFKL}}(10)
 \} \nonumber
\\
-\left(0-\frac{x_{02}}{x^2_{02}}\right)
\left(\frac{x_{12'}}{x^2_{12'}}-\frac{x_{02'}}{x^2_{02'}}\right)
\{
-M_{lin}(10|12')\left(N_{\mathtt{BFKL}}(2'0)+N_{\mathtt{BFKL}}(12')+N_{\mathtt{BFKL}}(10)\right)
\nonumber
 \\
 +\frac{1}{2}M^2_{lin}(10|12')+N_{\mathtt{BFKL}}(10)N_{\mathtt{BFKL}}(2'0)+N_{\mathtt{BFKL}}(10)N_{\mathtt{BFKL}}(12')
 \} \nonumber
\\
-\left(\frac{x_{12}}{x^2_{12}}-0\right)
\left(\frac{x_{12'}}{x^2_{12'}}-\frac{x_{02'}}{x^2_{02'}}\right)
\{
-M_{lin}(10|2'0)\left(N_{\mathtt{BFKL}}(12')+N_{\mathtt{BFKL}}(2'0)+N_{\mathtt{BFKL}}(10)\right)
\nonumber
 \\
 +\frac{1}{2}M^2_{lin}(10|2'0)+N_{\mathtt{BFKL}}(10)N_{\mathtt{BFKL}}(2'0)+N_{\mathtt{BFKL}}(10)N_{\mathtt{BFKL}}(12')
 \} \nonumber
\end{eqnarray}
\begin{eqnarray}\label{agkV3}
  &&+\left(\frac{x_{12}}{x^2_{12}}\frac{x_{12'}}{x^2_{12'}}+\frac{x_{02}}{x^2_{02}}\frac{x_{02'}}{x^2_{02'}}\right)
\left\{\frac{1}{2}M^2_{lin}(10|10)-M_{lin}(10|10)2N_{\mathtt{BFKL}}(10)+N^2_{\mathtt{BFKL}}(10)\right\}
\nonumber
\\
&&-\left(\frac{x_{12}}{x^2_{12}}\frac{x_{02'}}{x^2_{02'}}
+\frac{x_{02}}{x^2_{02}}\frac{x_{12'}}{x^2_{12'}}\right)
N^2_{\mathtt{BFKL}}(10) \nonumber
\end{eqnarray}
We read out the double cut contribution from  Eqs.~(\ref{agkV1})-(\ref{agkV3})
\begin{eqnarray}\label{doubleV1}
 &&\frac{d\sigma_{2P(\emph{double})}^{q\bar{q}}(x_{1},x_0)}{dY_0 d^2k}=
\frac{\bar{\alpha}_s}{2\pi}\frac{1}{(2\pi)^2}
\int d^2 x_2 \; d^2 x_{2'}e^{-ik(x_2-x_{2'})}
\\
&&\left(\frac{x_{12}}{x^2_{12}}\frac{x_{12'}}{x^2_{12'}}
 \frac{1}{2}\left\{
(M^2_{lin}(12|12')+M^2_{lin}(20|2'0))^2-M^2_{lin}(10|2'0)-M^2_{lin}(20|10)+M^2_{lin}(10|10) \right\} \hspace{0.7cm}
  \right. \nonumber
\\
&&\left.+ \frac{x_{02}}{x^2_{02}}\frac{x_{02'}}{x^2_{02'}}
 \frac{1}{2}\left\{
(M^2_{lin}(12|12')+M^2_{lin}(20|2'0))^2-M^2_{lin}(10|12')-M^2_{lin}(12|10)+M^2_{lin}(10|10) \right\} \hspace{0.7cm}
  \right. \nonumber
\\
&&\left.- \frac{x_{12}}{x^2_{12}}\frac{x_{02'}}{x^2_{02'}}
 \frac{1}{2}\left\{
(M^2_{lin}(12|12')+M^2_{lin}(20|2'0))^2-M^2_{lin}(10|2'0)-M^2_{lin}(12|10)\right\} \hspace{0.7cm}
  \right. \nonumber
\\
&&\left.+ \frac{x_{02}}{x^2_{02}}\frac{x_{12'}}{x^2_{12'}}
 \frac{1}{2}\left\{
(M^2_{lin}(12|12')+M^2_{lin}(20|2'0))^2-M^2_{lin}(10|12')-M^2_{lin}(20|10) \right\} 
  \right)\nonumber
\end{eqnarray}  
It is a tedious, but a straightforward calculation to find the single cut contribution in the form of 
\begin{eqnarray}\label{singleV1}
 &&\frac{d\sigma_{2P(\emph{single})}^{q\bar{q}}(x_{1},x_0)}{dY_0 d^2k}=
\frac{\bar{\alpha}_s}{2\pi}\frac{1}{(2\pi)^2}
\int d^2 x_2 \; d^2 x_{2'}e^{-ik(x_2-x_{2'})}
\\
&&\left(
\frac{x_{12}}{x^2_{12}}\frac{x_{12'}}{x^2_{12'}}
 \left\{
-\left(M_{lin}(12|12')+M_{lin}(20|2'0)\right)
\left(N_{\mathtt{BFKL}}(12)+N_{\mathtt{BFKL}}(20)
+N_{\mathtt{BFKL}}(12')+N_{\mathtt{BFKL}}(2'0)\right)
\right.\right. \nonumber
\\
&&\left. \left.
 +M_{lin}(20|10)\left(N_{\mathtt{BFKL}}(10)+N_{\mathtt{BFKL}}(12)+N_{\mathtt{BFKL}}(20)\right)
+M_{lin}(10|2'0)\left(N_{\mathtt{BFKL}}(10)+N_{\mathtt{BFKL}}(12')+N_{\mathtt{BFKL}}(2'0)\right)
\right.\right. \nonumber
\\
&&\left. \left.
-2M_{lin}(10|10)N_{\mathtt{BFKL}}(10)
\right\}
\right. \nonumber
\\
&&\left.
-\frac{x_{12}}{x^2_{12}}\frac{x_{02'}}{x^2_{02'}}
 \left\{
-\left(M_{lin}(12|12')+M_{lin}(20|2'0)\right)
\left(N_{\mathtt{BFKL}}(12)+N_{\mathtt{BFKL}}(20)
+N_{\mathtt{BFKL}}(12')+N_{\mathtt{BFKL}}(2'0)\right)
\right.\right. \nonumber
\\
&&\left. \left.
 +M_{lin}(12|10)\left(N_{\mathtt{BFKL}}(10)+N_{\mathtt{BFKL}}(12)+N_{\mathtt{BFKL}}(20)\right)
+M_{lin}(10|2'0)\left(N_{\mathtt{BFKL}}(10)+N_{\mathtt{BFKL}}(12')+N_{\mathtt{BFKL}}(2'0)\right)
\right\}
\right. \nonumber
\\
&&\left.
-\frac{x_{02}}{x^2_{02}}\frac{x_{12'}}{x^2_{12'}}
 \left\{
-\left(M_{lin}(12|12')+M_{lin}(20|2'0)\right)
\left(N_{\mathtt{BFKL}}(12)+N_{\mathtt{BFKL}}(20)
+N_{\mathtt{BFKL}}(12')+N_{\mathtt{BFKL}}(2'0)\right)
\right.\right. \nonumber
\\
&&\left. \left.
 +M_{lin}(20|10)\left(N_{\mathtt{BFKL}}(10)+N_{\mathtt{BFKL}}(12)+N_{\mathtt{BFKL}}(20)\right)
+M_{lin}(10|12')\left(N_{\mathtt{BFKL}}(10)+N_{\mathtt{BFKL}}(12')+N_{\mathtt{BFKL}}(2'0)\right)
\right\}
\right. \nonumber
\\
&&\left.
+\frac{x_{02}}{x^2_{02}}\frac{x_{02'}}{x^2_{02'}}
 \left\{
-\left(M_{lin}(12|12')+M_{lin}(20|2'0)\right)
\left(N_{\mathtt{BFKL}}(12)+N_{\mathtt{BFKL}}(20)
+N_{\mathtt{BFKL}}(12')+N_{\mathtt{BFKL}}(2'0)\right)
\right.\right. \nonumber
\\
&&\left. \left.
 +M_{lin}(12|10)\left(N_{\mathtt{BFKL}}(10)+N_{\mathtt{BFKL}}(12)+N_{\mathtt{BFKL}}(20)\right)
+M_{lin}(10|12')\left(N_{\mathtt{BFKL}}(10)+N_{\mathtt{BFKL}}(12')+N_{\mathtt{BFKL}}(2'0)\right)
\right.\right. \nonumber
\\
&&\left. \left.
-2M_{lin}(10|10)N_{\mathtt{BFKL}}(10)
\right\}
  \frac{}{} \right) \nonumber
\end{eqnarray}

Finally, the diffractive cut gives 
\begin{eqnarray}\label{diffV1}
 &&\frac{d\sigma_{2P(\emph{diff})}^{q\bar{q}}(x_{1},x_0)}{dY_0 d^2k}=
\frac{\bar{\alpha}_s}{2\pi}\frac{1}{(2\pi)^2}
\int d^2 x_2 \; d^2 x_{2'}e^{-ik(x_2-x_{2'})}\left(
\frac{x_{12}}{x^2_{12}}-\frac{x_{02}}{x^2_{02}}\right)\left(\frac{x_{12'}}{x^2_{12'}}-\frac{x_{02'}}{x^2_{02'}}\right)
\\
&&
\left\{N_{\mathtt{BFKL}}(10)-N_{\mathtt{BFKL}}(12)-N_{\mathtt{BFKL}}(20)\right\}
\left\{N_{\mathtt{BFKL}}(10)-N_{\mathtt{BFKL}}(12')-N_{\mathtt{BFKL}}(2'0)\right\}
\nonumber
\end{eqnarray} 

The contributions from the different cuts can be further simplified using the unitarity condition 
\eq{unitGen}, but the expressions become cumbersome  and we want to consider only a special case of 
$\frac{x_{12}}{x^2_{12}}\frac{x_{12'}}{x^2_{12'}}$ dipole splitting for illustrating the general
conclusion regarding the AGK violation. 
The cuts are:
\begin{itemize}
\item  \textbf{double cut} 
\end{itemize}
\begin{eqnarray}\label{doublex12}
\frac{1}{2}(4 N^2_{\mathtt{BFKL}}(10) - (N_{\mathtt{BFKL}}(10) - N_{\mathtt{BFKL}}(12) + N_{\mathtt{BFKL}}(20))^2
 - (N_{\mathtt{BFKL}}(10) - N_{\mathtt{BFKL}}(12') + N_{\mathtt{BFKL}}(2'0))^2 
\\ 
+ (N_{\mathtt{BFKL}}(12) + N_{\mathtt{BFKL}}(12') + N_{\mathtt{BFKL}}(20) - 2N_{\mathtt{BFKL}}(22') + N_{\mathtt{BFKL}}(2'0))^2) \nonumber
\end{eqnarray} 
\begin{itemize}
\item  \textbf{single cut} 
\end{itemize}
\begin{eqnarray}\label{singlex12}
-2(N^2_{\mathtt{BFKL}}(10) + 
        N^2_{\mathtt{BFKL}}(12) + (N_{\mathtt{BFKL}}(12') + N_{\mathtt{BFKL}}(20))(N_{\mathtt{BFKL}}(12') -N_{\mathtt{BFKL}}(22'))\\ + (N_{\mathtt{BFKL}}(12') +  N_{\mathtt{BFKL}}(20) - N_{\mathtt{BFKL}}(22')) N_{\mathtt{BFKL}}(2'0)
 - N_{\mathtt{BFKL}}(10)(N_{\mathtt{BFKL}}(20) + N_{\mathtt{BFKL}}(2'0))
\nonumber\\ + 
        N_{\mathtt{BFKL}}(12)(N_{\mathtt{BFKL}}(12') + N_{\mathtt{BFKL}}(20) - N_{\mathtt{BFKL}}(22') + N_{\mathtt{BFKL}}(2'0)) \nonumber
\end{eqnarray} 
\begin{itemize}
\item  \textbf{diffractive cut}
\end{itemize}~\begin{eqnarray}\label{diffx12}
\left\{N_{\mathtt{BFKL}}(10)-N_{\mathtt{BFKL}}(12)-N_{\mathtt{BFKL}}(20)\right\}
\left\{N_{\mathtt{BFKL}}(10)-N_{\mathtt{BFKL}}(12')-N_{\mathtt{BFKL}}(2'0)\right\}
\end{eqnarray}
It is clearly seen from Eqs.~(\ref{doublex12})-(\ref{diffx12}) that already in the single inclusive cross section
the original AGK cutting rules are violated in the sense that the contributions from the different cuts are not proportional to each other. This should be compared to the cuts contributing to the total cross section given in 
\eq{cutsDone}, which are proportional to the total cross section with coefficients $2\div-4\div1$.

It should be emphasized that, in contrary to the original AGK cutting rules, non of the cuts is expressed through
the total cross section as we show in Appendix~\ref{sec:R}. 

In the production of another gluon from one of the lower Pomerons (see \fig{fig:process2} ) 
\FIGURE[h]{ \centerline{
 \epsfig{file=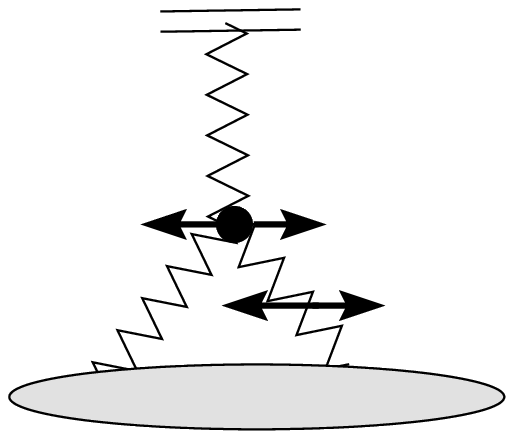,width=50mm} } \caption{
 The emission from the vertex in one to two Pomerons splitting.}
\label{fig:process2}}
one would expect a cancellation naively applying AGK cutting rules as shown \fig{fig:agk-cancel}. 
This does not happen if the gluon production is preceded by the emission from the vertex, because in this case 
the double cut \eq{doublex12} is not proportional to the single cut \eq{singlex12}. 

Another important remark is to be made, in the real gluon production from the vertex we took into account  only diagrams $A$, $B$, $B^*$
 and $D$  in \fig{fig:real-virt}.
 As we showed in Section~\ref{sec:agk} the cuts of diagram $A$ fully satisfied the AGK rules for the total cross section, and we expect the AGK violation to originate only from diagram $B$, $B^*$
 and $D$ in \fig{fig:real-virt}. This is not the case, because of the fact that the produced gluon
transverse coordinate in the amplitude $x_2$ differs from its coordinate $x_{2'}$ in the conjugate amplitude.
The cuts of diagram $A$ in \fig{fig:real-virt} for the dipole splitting $\frac{x_{12}}{x^2_{12}}\frac{x_{12'}}{x^2_{12'}}$ are given by 
\small
\begin{itemize}
\item  \textbf{double cut of diagram $A$} 
\end{itemize}
\vspace{-0.5cm}
\begin{eqnarray}\label{doubleAx12}
\frac{1}{2}\left(N_{\mathtt{BFKL}}(12) + N_{\mathtt{BFKL}}(12') + N_{\mathtt{BFKL}}(20) +N_{\mathtt{BFKL}}(2'0)-
 2N_{\mathtt{BFKL}}(22')\right)^2
\end{eqnarray}
\begin{itemize}
\item  \textbf{single cut of diagram $A$} 
\end{itemize}
\vspace{-0.5cm}
\begin{eqnarray}\label{singleAx12}
-\left(N_{\mathtt{BFKL}}(12) + N_{\mathtt{BFKL}}(12') + N_{\mathtt{BFKL}}(20) +N_{\mathtt{BFKL}}(2'0)\right)
\left(N_{\mathtt{BFKL}}(12) + N_{\mathtt{BFKL}}(12') + N_{\mathtt{BFKL}}(20) +N_{\mathtt{BFKL}}(2'0)-
 2N_{\mathtt{BFKL}}(22')\right)
\end{eqnarray}
\begin{itemize}
\item  \textbf{diffractive cut of diagram $A$}
\end{itemize}
\vspace{-0.5cm}
\begin{eqnarray}\label{diffAx12}
+\left(N_{\mathtt{BFKL}}(12)+ N_{\mathtt{BFKL}}(20) \right)
\left( N_{\mathtt{BFKL}}(12') +N_{\mathtt{BFKL}}(2'0)
\right)
\end{eqnarray}
\normalsize
The double \eq{doubleAx12}, the single \eq{singleAx12} and the diffractive  \eq{diffAx12} cuts
are proportional to each other only for $x_2=x_{2'}$. 

\eq{doublex12} - \eq{diffx12} does not allow us to calculate the double inclusive cross section
but we can easily to calculate the following observable
$ \int d^2 k_1 d^2 k_2 \int^{y_1}_0 \,d y_2 d^6  \sigma/ d^2 k_1 d^2 k_2 d y_1 d y_2$ which gives the multiplicity of particles with rapidity less than $y_1$ which accompany the produced jet at rapidity $y_1$ integrated over possible transverse momentum. From \eq{doublex12} - \eq{singlex12} one can conclude that
\bea
&&\int d^2 k_1 d^2 k_2 \int^{y_1}_0 \,d y_2\,\,\frac{ d^6  \sigma}{d^2 k_1 d^2 k_2 d y_1 d y_2} \,\,=\,\,\label{DINCM}\\
&&\frac{\bar{\alpha}_S}{2\pi}\frac{1}{2\pi}
\int d^2 x_2 \; \frac{x^2_{01}}{x^2_{12}\,x^2_{02}}\,4\,N_{\mathtt{BFKL}}(02) \,\Lb N_{\mathtt{BFKL}}(01) \,-\,
N_{\mathtt{BFKL}}(02)\,+\,N_{\mathtt{BFKL}}(12)\Rb\,\langle|n (y_1 - 0)|\rangle_P \notag
\eea
where $\langle|n (y_1 - 0)|\rangle_P$ is the average multiplicity in the rapidity window $y_1 - 0$.

One can see that cross section of \eq{DINCM} is not equal to zero and therefore, this equations confirms the main result of Ref. \cite{JalilianMarian:2004da}.

In the next Section we show processes of different multiplicity can be  extracted directly from the Glauber expression
for the single inclusive cross section.

\section{AGK rules in Glauber Formalism}\label{sec:agkglaub}
In this Section we use the Glauber expression for the single inclusive cross section including multiple rescatterings to find the contributions of different multiplicities to the total cross section. In the classical expression for the single inclusive cross section Eqs.~(\ref{Classline1})-(\ref{Classline4})  we included both elastic and inelastic rescatterings on a large target consisting of an infinite number of nucleons. In this Section 
we consider only two nucleons and show that in the leading order this coincides with the expression for two Pomeron 
exchange in Section~\ref{sec:agkinc}. In order to do this we need to separate  elastic and inelastic contributions in the Glauber expression Eqs.~(\ref{Classline1})-(\ref{Classline4}), since in its derivation we use cancellations of these two as follows. In Section~\ref{sec:noevol} we started by defining cross section 
$M_0(ij|ik)$, which includes both elastic and inelastic parts and was found by Glauber summation in Appendix~\ref{sec:B}.  In passing from Eqs.~(\ref{line1})-(\ref{line5}) to Eqs.~(\ref{Classline1})-(\ref{Classline4}) we used \eq{AINA1} obtained from 
 \eq{AINA10} with the help of \eq{AINXS}.  In this step we canceled some of the  inelastic terms with the elastic ones. Our goal is to reconstruct the classical expression for the single inclusive cross section before we make use  of these cancellations. To do this we just substitute the expression for the cross section $M_0(ij|ik)$ given by \eq{AINA10} in Eqs.~(\ref{line1})-(\ref{line5}).

After some lengthy calculations we obtain
\footnotesize
\begin{eqnarray}\label{lineIN1}
\frac{d\sigma^{q\bar{q} A \rightarrow GX}(x_1,x_0)}{d^2kdy}=\frac{\bar{\alpha_s}}{2\pi}\frac{1}{(2\pi)^2}\int d^2x_2 d^2{x}_{2'}e^{-ik(x_2-x_{2'})}\left\{\left(\frac{x_{12}}{x^2_{12}}-\frac{x_{02}}{x^2_{02}}\right)\left(\frac{x_{12'}}{x^2_{12'}}-\frac{x_{02'}}{x^2_{02'}}\right)
\right.
\\
\left.
\left(
1 - e^{-\frac{1}{2}(\sigma^{BA}(12)+\sigma^{BA}(20))} -
e^{-\frac{1}{2}(\sigma^{BA}(12')+\sigma^{BA}(2'0))}
+e^{-\frac{1}{2}(\sigma^{BA}(12)+\sigma^{BA}(20)+\sigma^{BA}(12')+\sigma^{BA}(2'0))+\hat\sigma_{in}(12|12')
+\hat\sigma_{in}(20|2'0)} \
\right)
\right.  \;\;\;\;\;\; 
\end{eqnarray}
\begin{eqnarray}\label{lineIN2}
\left.
-\left(\frac{x_{12}}{x^2_{12}}-\frac{x_{02}}{x^2_{02}}\right)\left(0-\frac{x_{02'}}{x^2_{02'}}\right)
\left(
1 - e^{-\frac{1}{2}(\sigma^{BA}(12)+\sigma^{BA}(20))} -
e^{-\frac{1}{2}\sigma^{BA}(10)}
+e^{-\frac{1}{2}(\sigma^{BA}(12)+\sigma^{BA}(20)+\sigma^{BA}(10))+\hat\sigma_{in}(12|10)} 
\right)
\right.
\end{eqnarray}
\begin{eqnarray}
\left.
-\left(\frac{x_{12}}{x^2_{12}}-\frac{x_{02}}{x^2_{02}}\right)\left(\frac{x_{12'}}{x^2_{12'}}-0\right)
\left(
1 - e^{-\frac{1}{2}(\sigma^{BA}(12)+\sigma^{BA}(20))} -
e^{-\frac{1}{2}\sigma^{BA}(10)}
+e^{-\frac{1}{2}(\sigma^{BA}(12)+\sigma^{BA}(20)+\sigma^{BA}(10))+\hat\sigma_{in}(20|10)} 
\right) \hspace{0.5cm}
\right.  \nonumber
\end{eqnarray}
\begin{eqnarray}\label{lineIN3}
\left.
-\left(0-\frac{x_{02}}{x^2_{02}}\right)\left(\frac{x_{12'}}{x^2_{12'}}-\frac{x_{02'}}{x^2_{02'}}\right)
\left(
1 -
e^{-\frac{1}{2}\sigma^{BA}(10)}- e^{-\frac{1}{2}(\sigma^{BA}(12')+\sigma^{BA}(2'0))} 
+e^{-\frac{1}{2}(\sigma^{BA}(12')+\sigma^{BA}(2'0)+\sigma^{BA}(10))+\hat\sigma_{in}(10|12')} 
\right) \hspace{0.5cm}
\right.  \nonumber
\end{eqnarray}
\begin{eqnarray}
\left.
-\left(\frac{x_{12}}{x^2_{12}}-0\right)\left(\frac{x_{12'}}{x^2_{12'}}-\frac{x_{02'}}{x^2_{02'}}\right)
\left(
1 -
e^{-\frac{1}{2}\sigma^{BA}(10)}- e^{-\frac{1}{2}(\sigma^{BA}(12)+\sigma^{BA}(20))} 
+e^{-\frac{1}{2}(\sigma^{BA}(12')+\sigma^{BA}(2'0)+\sigma^{BA}(10))+\hat\sigma_{in}(10|2'0)} 
\right) \hspace{0.5cm}
\right.  \nonumber
\end{eqnarray}
\begin{eqnarray}\label{lineIN4}
+\left(\frac{x_{12}}{x^2_{12}}\frac{x_{12'}}{x^2_{12'}}+\frac{x_{02}}{x^2_{02}}\frac{x_{02'}}{x^2_{02'}}\right)
\left(
1 -
2\;e^{-\frac{1}{2}\sigma^{BA}(10)}+e^{-\frac{1}{2}2\sigma^{BA}(10)+\hat\sigma_{in}(10|10)} 
\right)
-\left(\frac{x_{12}}{x^2_{12}}\frac{x_{02'}}{x^2_{02'}}+\frac{x_{02}}{x^2_{02}}\frac{x_{12'}}{x^2_{12'}}\right)\left(1 - e^{-\frac{1}{2}\sigma^{BA}(10)}\right)^2 
\left. \frac{}{}
\right\}
\end{eqnarray} 
\normalsize
where, for simplicity, we absorbed the profile function $T(b;R_A)$ in the definition of the scattering cross sections $\sigma^{BA}$ and $\hat\sigma_{in}$. The same expression can be directly applying Glauber formula as shown in Appendix~\ref{sec:R}.

Now, we expand the exponentials Eqs.~(\ref{lineIN1})-(\ref{lineIN4}) 
in powers of $\hat\sigma^{in}$ to the second order to find the contributions that correspond to the double cut of two Pomeron exchange found in \eq{doubleV1}. After some algebra we obtain the same structure we have in \eq{doubleV1}. Expansion in powers of $(\sigma^{BA})^2$ reproduces the diffractive cut in \eq{doublex12}, and, finally, the expansion in powers of
$\hat\sigma^{in}\sigma^{BA}$ corresponds to the single cut in \eq{singleV1}. 
It is important to note, that the classical expression in Eqs.~(\ref{lineIN1})-(\ref{lineIN4}) does not account for the evolution effects, and thus should not reproduce  Eqs.~(\ref{lineIN1})-(\ref{lineIN4}) in more general case, when the evolution is switched on. The way to do this is to go back to Eqs.~(\ref{line1})-(\ref{line5}) and substitute solutions to the evolution equations $M(ij|ik)$ and $N(ij)$ in place of $M_0(ij|ik)$ and $N_0(ij)$ as described in Section~\ref{sec:evol}. Then, expand  to the desired order in $\hat\sigma_{in}$ and $\sigma_{BA}$ to obtain contributions of different multiplicities. However, because of the fact that $M(ij|ik)$ and $N(ij)$ are solutions to non-linear equations with "Pomeron splittings" where "cut Pomeron" can split to two "uncut Pomerons" at some rapidity, fixing the interaction type~($\hat\sigma_{in}$ or $\sigma_{BA}$)  by expanding the solution bring little or even no information about the multiplicity of the produced particles in the rapidity window between the target and the projectile. Thus, the proper way for studying the multiplicity distribution is to attach Pomerons~(cut or uncut) to a triple Pomeron vertex at some given rapidity, as we did deriving Eqs.~(\ref{agkV1})-(\ref{agkV3}).

\vspace{2cm}

\section{Conclusions}\label{sec:conclusions}
The central result of this paper is the proof of the validity of the Abramovsky-Gribov-Kancheli cutting rules in pQCD for the total cross section, and the demonstration of their violation for a particle production from the vertex. This violations happens already for one gluon production from the triple Pomeron vertex as shown by explicit calculation of contributions from different cuts to the inclusive cross section of two Pomeron exchange. The general conclusion can be formulated as follows:
AGK rules valid for 
\begin{itemize}
\item  total cross section;
\item  production of any number of gluons from Pomerons either before or after triple Pomeron vertex. 
\end{itemize}
AGK rules violated for 
\begin{itemize}
\item one and more gluon productions from the triple Pomeron vertex.
\end{itemize}
This result allows to build a  generating functional that will incorporate all multiparticle production processes in the spirit of Ref.\cite{Levin:2007yv}, and find the equation for diffractive dissociation of Ref.~\cite{KLDD}.

As a by product of our analysis we derived a generalized form of the Balitsky-Kovchegov evolution equation for a non-diagonal cross section of the dipole scattering for a dipole having different coordinates in the amplitude and the conjugate amplitude. This generalised BK equation is easily solved by a linear combination of the solutions to the BK equation. This particular form of the solution preserves the adjoint~(dipole) structure of the single inclusive cross section found by Kovchegov and Tuchin \cite{Kovchegov:2001sc} to any order in the evolution. 

The generalized BK equation found in the present paper is extremely useful for a systematic treatment of multigluon productions, multiparticle correlations including jet production in diffractive dissociation  and  multiplicity distributions in more general cases.

\section* {Acknowledgements}
We are grateful to Jochen Bartels, Sergey Bondarenko, Errol Gotsman, Yura Kovchegov,  Lev Lipatov, Cyrille Marquet, Uri Maor and Kirill Tuchin  for fruitful  discussions on the subject. Our special thanks go to Yura Kovchegov who found a mistake in the previous version of this paper which led us to  an incorrect  violation of Kovchegov-Tuchin formula for single inclusive cross section.
One of us (A.P.) thanks
Nestor Armesto and Carlos Pajares for their hospitality  at University of Santiago de Compostela during the work of the paper. A.P. is also thankful to the organizers of the Les Houches School on Hadronic collisions $2008$, in particular, to Yuri Dokshitzer, Francois Gelis and Edmond Iancu for their interest in the results of this study presented at the school.

This research was supported  in part by the Israel Science Foundation, founded by the Israeli Academy of Science
and Humanities, by BSF grant $\#$ 20004019 and by
a grant from Israel Ministry of Science, Culture and Sport and
the Foundation for Basic Research of the Russian Federation.

~

\newpage

\Large
\textbf{Appendices}
\normalsize

\appendix
\renewcommand{\theequation}{A-\arabic{equation}}
\setcounter{equation}{0}  

\begin{boldmath}
\section{Light Cone Perturbative QCD at high energy (simple diagrams)} \label{sec:A}
\end{boldmath}
In this Appendix we calculate simplest diagrams in the Light Cone Perturbation Theory~(LCPT) applied to QCD, in particular, we show how the minus signs leading to the real-virtual cancellations appear in this framework. The
rules for LCPT in QCD were formulated Ref.~\cite{BRLE}. LCPT is very useful for high energy scattering since it represents an intuitive space-time picture of the scattering process. As a simple example of LCPT calculations consider
a diagram where a soft gluon $k$ is emitted from the heavy quark, and then the heavy quark interacts with the target as shown in \fig{fig:quarkgluon}. The interaction with the target is mediated by the  gluon $m$  and the intermediate states are denoted by the vertical dotted lines with light-cone energy denominators $D_1$ and $D_2$ are calculated from  
\begin{eqnarray}\label{denoms}
D=\sum_i p^{i}_{-} - \sum_f p^{f}_{-} 
\end{eqnarray}
where $\sum_i p^{i}_{-}$ and $\sum_f p^{f}_{-}$ are the sums of the light-cone energies of the intermediate and the initial~(or equivalently final) states, correspondingly. 
The denominators $D_1$ and $D_2$ are readily found using \eq{denoms}

\begin{eqnarray}\label{D1}
D_1=(p-k)_{-}+k_{-} +P_{-}-p_{-} -P_{-}=\frac{(p-k)_{\perp}^2}{(p-k)_{+}}+\frac{k_{\perp}^2}{k_{+}}
+\frac{p_{\perp}^2}{p_{+}}\simeq\frac{k_{\perp}^2}{k_{+}} =k_{-}
\end{eqnarray}

\begin{eqnarray}\label{D2}
D_2=(p-k-m)_{-}+k_{-}+m_{-} +P_{-}-p_{-} -P_{-}=\frac{(p-k-m)_{\perp}^2}{(p-k-m)_{+}}+\frac{k_{\perp}^2}{k_{+}}+
\frac{m_{\perp}^2}{m_{+}}
+\frac{p_{\perp}^2}{p_{+}}\simeq\frac{k_{\perp}^2}{k_{+}} \simeq \frac{m_{\perp}^2}{m_{+}} =m_{-}
\end{eqnarray}
 where we used the fact that $p_{+}\gg k_{+} \gg m_{+}$ in accordance with the Regge kinematics.
\FIGURE[h]{
\centerline{\epsfig{file=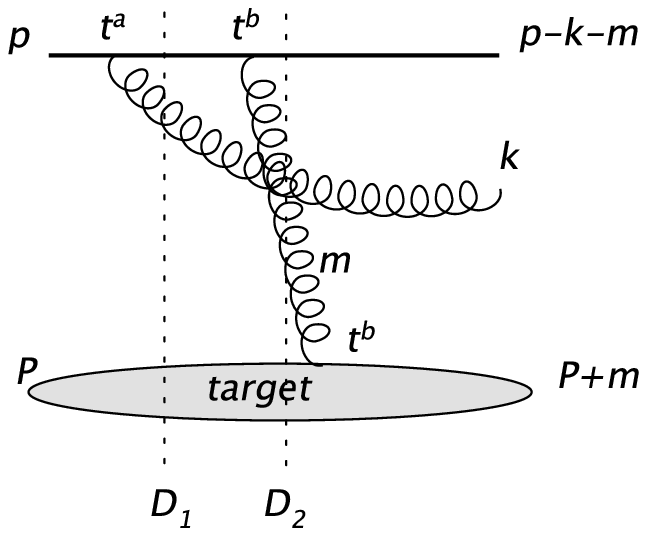,width=60mm}}
\caption{Emission of gluon in   Light Cone Perturbative QCD  . }
\label{fig:quarkgluon} }

The expression for the diagram depicted in \fig{fig:quarkgluon} is given by 
\begin{eqnarray}\label{diag1}
\frac{\bar{u}(P+m)}{\sqrt{(P+m)_+}}gt^b\gamma^{\lambda}\frac{u(P)}{\sqrt{(P)_+}}
\frac{1}{D_2}\frac{g^{\mu\lambda}}{m_{+}}\frac{\bar{u}(p-k-m)}{\sqrt{(p-k-m)_+}}gt^b\gamma^{\mu}\frac{u(p-k)}{\sqrt{(p-k)_+}}
\frac{1}{D_1}\frac{\bar{u}(p-k)}{\sqrt{(p-k)_+}}gt^a\gamma^{\nu}\frac{u(p)}{\sqrt{p_+}}
\end{eqnarray}
After some simplifications due to the normalization properties of the spinors in the kinematic regime  $p_{+}\gg k_{+} \gg m_{+}$ we obtain
\begin{eqnarray}\label{diag1a}
4g^2t^b\otimes t^b\frac{1}{m^2_{\perp}} 2gt^a\frac{ \epsilon_{-}}{k_{-}}=
4g^2t^b\otimes t^b\frac{1}{m^2_{\perp}} 2gt^a\frac{ k_{\perp}\cdot\epsilon_{\perp}}{k^2_{\perp}}
\end{eqnarray}
where we used the fact that $\epsilon\cdot k=0=\epsilon_{+}k_{-}+\epsilon_{-}k_{+}-\epsilon_{\perp}\cdot k_{\perp}\simeq \epsilon_{-}k_{+}-\epsilon_{\perp}\cdot k_{\perp}$.

Next, we calculate the so-called late emission diagram where the emission of the gluon $k$ occurs after the 
interaction as shown in \fig{fig:quarkgluon-late}
\FIGURE[h]{
\centerline{\epsfig{file=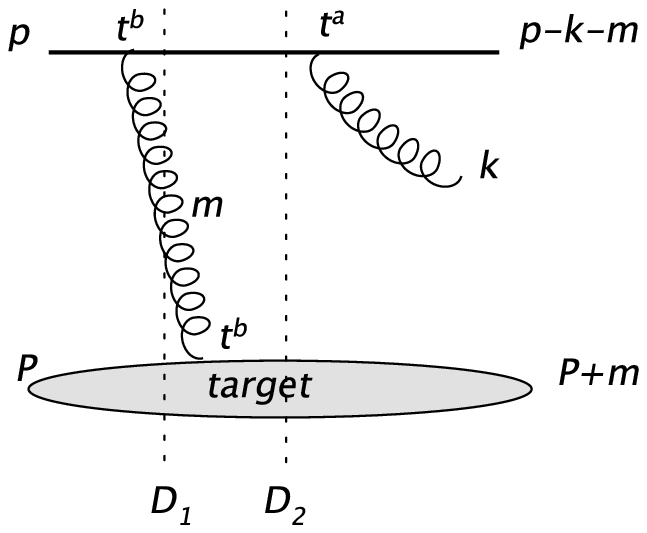,width=60mm}}
\caption{Late emission diagram in LCPT. This diagram is opposite in sign w.r.t the early emission in \fig{fig:quarkgluon} }
\label{fig:quarkgluon-late} }

As in the case of \fig{fig:quarkgluon} we write the light-cone energy denominators
\begin{eqnarray}\label{D1-late}
D_1=(p-m)_{-}+m_{-} +P_{-}-p_{-} -P_{-}=\frac{(p-m)_{\perp}^2}{(p-m)_{+}}+\frac{m_{\perp}^2}{m_{+}}
+\frac{p_{\perp}^2}{p_{+}}\simeq\frac{m_{\perp}^2}{m_{+}} =m_{-}
\end{eqnarray}

\begin{eqnarray}\label{D2-late}
D_2=(p-m)_{-}+m_{-} +(P+m)_{-}-(p-k-m)_{-} -(P+m)_{-}-k_{-}\simeq -k_{-}
\end{eqnarray}

In \eq{D2-late} we used the equivalence of the initial and the final light-cone energies. Note that due to the presence of the emitted gluon $k$ in the final state $D_2$ generates a minus sign.
The rest of the calculation is straightforward and in full analogy with \eq{diag1}. The final answer for the late 
emission diagram in \fig{fig:quarkgluon-late} is given by 
 \begin{eqnarray}\label{diag2a}
-4g^2t^b\otimes t^b\frac{1}{m^2_{\perp}} 2gt^a\frac{ \epsilon_{-}}{k_{-}}=
4g^2t^b\otimes t^b\frac{1}{m^2_{\perp}} 2gt^a\frac{ k_{\perp}\cdot\epsilon_{\perp}}{k^2_{\perp}}
\end{eqnarray}
and differs from \eq{diag1a} only by a minus sign coming from the light-cone energy denominator \eq{D2-late}.

Here we presented a simplified case only one quark scattered off the target. In general, any late emission diagram will generate a minus sign w.r.t early emissions.

\renewcommand{\theequation}{B-\arabic{equation}}
\setcounter{equation}{0}  

\begin{boldmath}
\section{Calculation of $M_0(il|kl)$} \label{sec:B}
\end{boldmath}

 In this Appendix  we derive the explicit expression for $M_0(il|kl)$ for dipole-nucleus scattering in the Glauber approach assuming the Born Approximation of perturbative QCD (the exchange of two Coulomb-like gluons)
for the  dipole-nucleon interaction.  It is well known that the Glauber formula
for the imaginary part of the scattering amplitude in this case has the following form
\beq \label{AGF1}
N_0(10)\,\,=\,\,1\,\,-\,\,e^{- \frac{1}{2}\sigma^{BA}(x^2_{10} )\,T\left(b;R_A\right)}
\eeq
with 
\beq \label{AXSBA}
 \sigma^{BA}(10)\,\,\equiv\,\,\sigma^{BA}(x_{10} )\,\,=\,\,2\,\bas\int\,\frac{d^2 l}{l^2}\,\left( 1\,\,-\,\,e^{- i l\cdot x_{10}}\right)\,IF(l)
\eeq
where $l$ is the transverse momentum of  one of two gluons in the Born Approximation diagram.

The function $IF(l)$ in \eq{AXSBA} denotes so-called impact factor and describes the interaction
 of two gluon with a nucleon which depends on $l^2$.
In our further calculations we do not need to know the explicit form of $IF(l)$.

The profile function $T\left(b;R_A\right)$ determines the number nucleons that can interact with the dipole at 
a given impact parameter $\vec{b} =\frac{1}{2} ( \vec{x}_1 + \vec{x}_0)$. 
The profile function $T\left(b;R_A\right)$ is defined through  the nucleon density in a nucleus $\rho (b,z)$ as
\beq 
T\left(b;R_A\right) \,\,=\,\,\int\,d z \,\rho (b,z);\,\,\,\,\,\,\,\,\,\int \,d^2 b\,\int^{+\infty}_{- \infty}  d z\,\,
\rho (b,z)\,\,=\,\,A
\eeq
 In the simple model
$\rho (b,z) \,=\,\rho_0\,\Theta \Lb R_A - b\Rb\, \Theta \Lb \sqrt{ R^2_A - b^2} - z\Rb $, where  $R_A$ is the nucleus radius.

 Let us consider a dipole having different sizes at the interaction time $\tau=0$ in the amplitude and conjugate amplitude, namely, a dipole with quark~(antiquark) coordinates $x_1$ ($x_2$) at $\tau=0$ in amplitude and $x_1$ ($x_{2'}$) in the conjugate amplitude. A more general case with all the coordinates being different is of no use in our derivation and thus will be omitted. The cross section $M(12|12')$ is normalized such  that  $M(12|12)\,=\,2\,N(12)$.

\FIGURE[ht]{
\centerline{\epsfig{file=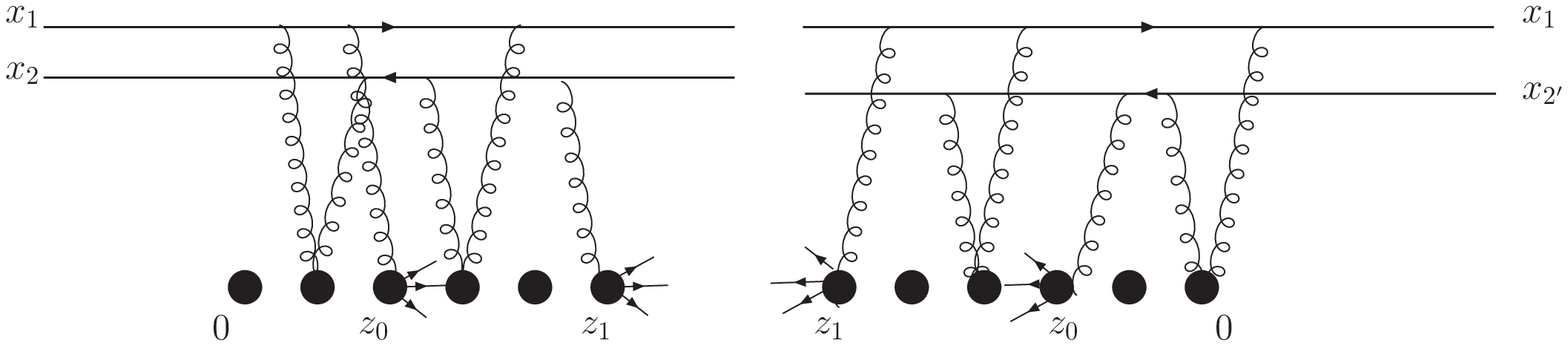,width=160mm}}
\caption{ Amplitude $M_0\Lb 12|12'\Rb$ in Glauber approach. } 
\label{dida} }
As one can see in \fig{dida} the interaction of the dipole $"12"$ as well as the dipole $"12'"$ with a nucleus  includes an additional elastic rescattering
as well as the inelastic interaction with the nucleons of the nucleus
at points $z_i$.

To include  both of the processes we  first write the formula that sums all possible inelastic interactions. It has the form of  \cite{LERY,KOMU}
\bea
&& M^{in}_0 \Lb 12|12'\Rb= \label{AMBA} \\
&& \times   \int^{2R_A(b)}_{0}\!\!\!\rho\, \hat\sigma_{in}(12|12')\,d
z_0\,e^{-\frac{1}{2}(\sigma^{BA}(12) +\sigma^{BA}(12'))\,\rho\,2\,R_A(b) }
\,\sum^{\infty}_{n=0} 
\int^{2 R_A(b)}_{z_0}\!\!d\,z_{1}\, \dots \int^{2 R_A(b)}_{z_{n -2}}\!\!
dz_{n-1}\int^{2 R_A(b)}_{z_{n-1} }\!\!d z_n\,
\rho^n\,\hat\sigma^n_{in}(12|12')\, \nonumber 
\eea

The factor of  $ \exp\Lb-\frac{1}{2}(\sigma(12) +\sigma(12'))\,\rho\,2\,R_A(b) \Rb$  in \eq{AMBA}describes the fact that there are no inelastic interactions between points $z_1, z_2 \dots z_{n}$.  
Any of the inelastic interactions occurs  with the same nucleon in amplitude and conjugated amplitude since the contributions of the inelastic processes with different nucleons can be expressed through the nucleon correlations in the  nucleus which are suppressed in heavy nuclei.
The sum in \eq{AMBA} exponentiates and gives 

\beq \label{AINA10}
 M^{in}_0 \Lb 12|12'\Rb\,\,=\left( e^{ +\hat\sigma_{in}(12|12')\,\,T\left(b;R_A\right) }\,\,\,-\,\,\,
1\right)e^{-\frac{1}{2}\left\{\sigma^{BA}(12) +\sigma^{BA}(12')\right\}\,T\left(b;R_A\right)}
\eeq
 
\bea 
\hat\sigma_{in}(12|12')\,\,&=&\,\,2\,\int \frac{d^2 l}{l^2}\,\left( 1\,\,-\,\,e^{- i l\cdot x_{12}}\right)\,\times \,\left( 1\,\,-\,\,e^{- i l\cdot x_{12'}}\right)\,IF(l) \notag \\
   &=& \frac{1}{2}\,\Lb \sigma^{BA}(12) + \sigma^{BA}(12')\,\,-\,\sigma^{BA}(22') \Rb\label{AINXS} \\
 &\mbox{where} & \sigma^{BA}(22')\,\,=\,\,\sigma^{BA}(| \vec{x}_{12} - \vec{x}_{12'}|)\,\,\label{INXS}
\eea
 
Using \eq{AINXS} and \eq{INXS}  we can reduce \eq{AINA10} to
\beq \label{AINA1}
 M^{in}_0 \Lb 12|12'\Rb\,\,=\,\, e^{ - \frac{1}{2}\,\sigma^{BA}(22')\,\,T\left(b;R_A\right) }\,\,\,-\,\,\,
e^{-\frac{1}{2}\left\{\sigma^{BA}(12) +\sigma^{BA}(12')\right\}\,T\left(b;R_A\right)}
\eeq
To obtain the final formula we need to add to \eq{AINA1} the contribution of the elastic scattering which has the obvious form
\beq \label{AEL}
M^{el}_0 \Lb 12|12'\Rb\,\,=\,\,\Lb 1\,-\,e^{ - \frac{1}{2}\,\sigma^{BA}(12)\,\,T\left(b;R_A\right)}\,\Rb
\times\,\Lb 1\,-\,e^{- \frac{1}{2}\,\sigma^{BA}(12')\,\,T\left(b;R_A\right)} \Rb
\eeq

Following the lines of Refs.~\cite{Kovchegov:2001ni,K} we use the explicit expression for the powers of the exponentials in \eq{AINA1} and \eq{AEL}, namely,
\begin{eqnarray} \label{power}
\frac{1}{2}\,\sigma^{BA}(12)\,\,T\left(b;R_A\right)=\frac{1}{4}x^2_{12}Q^2_{s}
\end{eqnarray}
where $Q_s$ is the saturation scale used in the McLerran-Venugopalan model \cite{MV,LERY,LERY1} defined as
\begin{eqnarray}\label{sat}
Q^2_{s}(b)=2\alpha_s \rho T(b)
\end{eqnarray}

Using this simple notation we can readily find from \eq{AINA1} and \eq{AEL}

\bea\label{MnotA}
M_0(12|12') \,\,&=&\,\,\,1+e^{-x^2_{22'}Q^2_{s}/4}-e^{-x^2_{12}Q^2_{s}/4}-e^{-x^2_{12'}Q^2_{s}/4} 
\eea
\bea\label{MnotB}
M_0(12|10) &=&1+e^{-x^2_{20}Q^2_{s}/4}-e^{-x^2_{12}Q^2_{s}/4}-e^{-x^2_{10}Q^2_{s}/4}  
\eea
\bea\label{MnotD} 
M_0(10|10) &=& 2(1-e^{-x^2_{10}Q^2_{s0}/4}) 
\eea
 and  other required combinations.


 \renewcommand{\theequation}{C-\arabic{equation}}
\setcounter{equation}{0}  

\begin{boldmath}
\section{First step of the evolution in the Glauber approach} \label{sec:D}
\end{boldmath}
In this section we consider the first step of the evolution for the dipole $"12"("12'")$ having different 
transverse coordinates in the amplitude and the conjugate amplitude, namely, we emit one extra soft gluon $"3"$ as depicted in \fig{fig:22tag3}. The evolution in this case is described by the equation
 Eqs.~(\ref{evolMA1}) -(\ref{evolMA6}) and its first iteration gives the required  emission 
of soft gluon $"3"$. We analyze this situation both using Eqs.~(\ref{evolMA1}) -(\ref{evolMA6}) and show 
its equivalence to the direct application of the Glauber theory.

The soft gluon $"3"$ can be emitted before or after the interaction time $\tau=0$. Firstly, we consider  the contribution of the diagram A in \fig{fig:22tag3} where the soft gluon is emitted before $\tau=0$ in both the amplitude and the conjugate amplitude. The direct application of the Glauber expressions for the scattering cross section of two dipoles $"13"$ and  $"32"("32'")$ found in Appendix~\ref{sec:B} gives in this case
\begin{eqnarray}\label{glaubA}
&&\mbox{\tiny elastic scattering in the amplitude and conjugated amplitude} 
\hspace{2cm}
\mbox{\tiny inelastic scattering in the amplitude and conjugated amplitude}
\nonumber\\
&& \overbrace{\left(1-e^{-(x^2_{13}+x^2_{32})Q^2_s/4}\right)\left(1-e^{-(x^2_{13}+x^2_{32'})Q^2_s/4}\right)}
\hspace{1cm}
+
\hspace{1cm}
\overbrace{e^{-x^2_{22'}Q^2_s/4}-e^{-(2x^2_{13}+x^2_{32}+x^2_{32'})Q^2_s/4}}\vspace{1cm}\nonumber\\
&& =1+e^{-x^2_{22'}Q^2_s/4}-e^{-(x^2_{13}+x^2_{32})Q^2_s/4}-e^{-(x^2_{13}+x^2_{32'})Q^2_s/4}
\end{eqnarray}
times the Kernel of the dipole splitting (the emission soft gluon $"3"$) given by 
\begin{eqnarray}\label{KernelA}
\frac{\bar{\alpha_s}}{2\pi}\left(\frac{x_{13}}{x^2_{13}}-\frac{x_{23}}{x^2_{23}}\right)\left(\frac{x_{13}}{x^2_{13}}-\frac{x_{2'3}}{x^2_{2'3}}\right)
\end{eqnarray}
On the other hand we can use \eq{evolMA2} and \eq{evolMA3} to obtain the same result as follows

\begin{eqnarray}\label{glaubEqA}
&& M_0(13|13)+M_0(32|32')
+M_0(13|13)M_0(32|32')-2 M_0(32|32')N_0(13)
\nonumber
\\&&-M_0(13|13)\left\{N_0(32)+N_0(32')\right\}+N_0(13)\left\{N_0(32')+N_0(32)\right\}\\
&&=2\left(1-e^{-x^2_{13}Q^2_s/4}\right)+\left(1+e^{-x^2_{22'}Q^2_s/4}-e^{-x^2_{32}Q^2_s/4}-e^{-x^2_{32'}Q^2_s/4}\right)\nonumber \\
&& 2\left(1-e^{-x^2_{13}Q^2_s/4}\right)\left(1+e^{-x^2_{22'}Q^2_s/4}-e^{-x^2_{32}Q^2_s/4}-e^{-x^2_{32'}Q^2_s/4}\right) \nonumber \\
&&-2\left(1+e^{-x^2_{22'}Q^2_s/4}-e^{-x^2_{32}Q^2_s/4}-e^{-x^2_{32'}Q^2_s/4}\right)\left(1-e^{-x^2_{13}Q^2_s/4}\right) 
 \nonumber \\
&& -2\left(1-e^{-x^2_{13}Q^2_s/4}\right)\left(1-e^{-x^2_{32}Q^2_s/4}+1-e^{-x^2_{32'}Q^2_s/4}\right)
\nonumber \\
&& +\left(1-e^{-x^2_{13}Q^2_s/4}\right)\left(1-e^{-x^2_{32}Q^2_s/4}+1-e^{-x^2_{32'}Q^2_s/4}\right)
\nonumber \\
&& =2\left(1-e^{-x^2_{13}Q^2_s/4}\right)+\left(1+e^{-x^2_{22'}Q^2_s/4}-e^{-x^2_{32}Q^2_s/4}-e^{-x^2_{32'}Q^2_s/4}\right)
\nonumber \\
&& -\left(1-e^{-x^2_{13}Q^2_s/4}\right)\left(1-e^{-x^2_{32}Q^2_s/4}+1-e^{-x^2_{32'}Q^2_s/4}\right)
\nonumber \\
&&=1+e^{-x^2_{22'}Q^2_s/4}-e^{-(x^2_{13}+x^2_{32})Q^2_s/4}-e^{-(x^2_{13}+x^2_{32'})Q^2_s/4} \nonumber
\end{eqnarray}
In the limit of $x_{2}=x_{2'}$ this equals to the total cross section of the two dipole scattering 
$2(1-e^{-(x^2_{13}+x^2_{32})Q^2_s/4})$ as expected from the optical theorem.

Next, we want to obtain the contribution of diagram~B in \fig{fig:22tag3}. The direct application of the Glauber formulae gives 
\begin{eqnarray}\label{glaubB}
&&\mbox{\tiny elastic scattering in the amplitude and conjugated amplitude} 
\hspace{2cm}
\mbox{\tiny inelastic scattering in the amplitude and conjugated amplitude}
\nonumber\\
&& \overbrace{\left(1+e^{-(x^2_{13}+x^2_{32})Q^2_s/4}\right)\left(1+e^{-x^2_{12'}Q^2_s/4}\right)}
\hspace{1cm}
+
\hspace{1cm}
\overbrace{e^{-(x^2_{32}+x^2_{32'})Q^2_s/4}-e^{-(2x^2_{13}+x^2_{32}+x^2_{32'})Q^2_s/4}}\vspace{1cm}\\
&& =1+e^{-(x^2_{32}+x_{32'})Q^2_s/4}-e^{-(x^2_{13}+x^2_{32})Q^2_s/4}-e^{-x^2_{12'}Q^2_s/4} \nonumber
\end{eqnarray}
 
The same result one obtains using \eq{evolMA4} in the evolution equation for $M(12|12')$ and the eikonal formulae for one dipole scattering
\begin{eqnarray}\label{glaubEqB}
&& M_0(13|12')\left\{1-N_0(32)\right\}+N_0(12')N_0(32) \\
&&=\left(1+e^{-x^2_{32'}Q^2_s/4}-e^{-x^2_{13}Q^2_s/4}-e^{-x^2_{12'}Q^2_s/4}\right)\left(1-1+e^{-x^2_{32}Q^2_s/4}\right)+\left(1-e^{-x^2_{12'}Q^2_s/4}\right)\left(1-e^{-x^2_{32}Q^2_s/4}\right)\nonumber
\\
&& =1+e^{-(x^2_{32}+x_{32'})Q^2_s/4}-e^{-(x^2_{13}+x^2_{32})Q^2_s/4}-e^{-x^2_{12'}Q^2_s/4} \nonumber
\end{eqnarray}
The contribution from diagram~$B^*$ is found in analogous way.

The remaining diagrams~$R$, $R^*$, $D$, $E$ and $F$ change the cross section merely by multiplication of the Kernel of the emission of the soft gluon $"3"$, in other words, when gluon $"3"$ is emitted in those diagrams, only the initial dipole interacts with the target as 

\begin{eqnarray}\label{glaubB1}
M_0(12|12')=1+e^{-x^2_{22'}Q^2_s/4}-e^{-x^2_{12}Q^2_s/4}-e^{-x^2_{12'}Q^2_s/4} 
\end{eqnarray}
 
In this way we show that evolution equation Eqs.~(\ref{evolMA1simple})-(\ref{evolMA4simple}), correctly describe the evolution to the first order. The higher order iterations can be found in the same way.


 \renewcommand{\theequation}{D-\arabic{equation}}
\setcounter{equation}{0}  

\begin{boldmath}
\section{Glauber expression for single inclusive cross section: AGK cuts} \label{sec:R}
\end{boldmath}
In this Appendix we show how the Glauber expression for the single inclusive cross section with inelastic part retained found in Section~\ref{sec:agkglaub}~(see Eqs.~(\ref{lineIN1})-(\ref{lineIN4})) can be obtained directly applying the Glauber formula. The cross section for single inclusive production consists of terms coming from diagram $A$ in \fig{fig:real-virt}(see \eq{lineIN2}), diagrams $B$ and $B^*$~(see \eq{lineIN3}) and diagram $D$~(see \eq{lineIN4}). 
We follow the lines of the Appendix~\ref{sec:D} and readily write the Glauber expression for diagram $A$ in \fig{fig:real-virt} as 
\small
\begin{eqnarray}\label{glaubAinel}
\small
&& \hspace{2cm}\mbox{\tiny elastic scattering in the amplitude and conjugated amplitude} 
\nonumber\\
&& \overbrace{\left(1-e^{-\frac{1}{2}\{\sigma^{BA}(12)+\sigma^{BA}(20)\}}\right)
\left(1-e^{-\frac{1}{2}\{\sigma^{BA}(12')+\sigma^{BA}(2'0)\}}\right)
}
\hspace{1cm}
\\
&&\hspace{2cm}\mbox{\tiny inelastic scattering in the amplitude and conjugated amplitude}  \nonumber\\
&&+
\overbrace{e^{-\frac{1}{2}\{\sigma^{BA}(12)+\sigma^{BA}(20)+\sigma^{BA}(12')+\sigma^{BA}(2'0)\}}
-e^{-\frac{1}{2}\{\sigma^{BA}(12)+\sigma^{BA}(20)+\sigma^{BA}(12')+\sigma^{BA}(2'0)\}
+\hat\sigma_{in}(12|12')+\hat\sigma_{in}(20|2'0)}}\vspace{1cm} \nonumber\\
&& =1-e^{-\frac{1}{2}\{\sigma^{BA}(12)+\sigma^{BA}(20)\}}-e^{-\frac{1}{2}\{\sigma^{BA}(12')+\sigma^{BA}(2'0)\}}
\nonumber
\\
&& \hspace{0.5cm}-e^{-\frac{1}{2}\{\sigma^{BA}(12)+\sigma^{BA}(20)+\sigma^{BA}(12')+\sigma^{BA}(2'0)\}
+\hat\sigma_{in}(12|12')+\hat\sigma_{in}(20|2'0)} \nonumber
\end{eqnarray}
\normalsize
where we kept $2\neq2'$~(coordinate of gluon in the amplitude and the conjugate amplitude) because we fix the momentum of the emitted gluon, and absorbed $T(b;R_A)$ in the definition of $\sigma^{BA}$.
In the diagram~$A$ in \fig{fig:real-virt} we depicted only  $\frac{x_{12}}{x^2_{12}}\frac{x_{12'}}{x^2_{12'}}$
splitting, all other kernel have the same Glauber expression. This is not the case as far diagrams of the type  $B$, $B^*$ and $D$ are concerned, their Glauber expression  depend on the dipole splitting, since some inelastic contributions are suppressed in large $N_C$ limit. We have already discussed this point deriving the single inclusive cross section Eqs.(\ref{line1})-(\ref{line5}) in Section~\ref{sec:noevol}.
The diagram $B$, as it is  shown in \fig{fig:real-virt}, brings
\begin{eqnarray}\label{glaubBinel}
\small
&& \hspace{2cm}\mbox{\tiny elastic scattering in the amplitude and conjugated amplitude} 
\nonumber\\
&& \overbrace{\left(1-e^{-\frac{1}{2}\{\sigma^{BA}(12)+\sigma^{BA}(20)\}}\right)
\left(1-e^{-\frac{1}{2}\sigma^{BA}(10)}\right)
}
\hspace{1cm}
\\
&&\hspace{2cm}\mbox{\tiny inelastic scattering in the amplitude and conjugated amplitude}  \nonumber\\
&&+
\overbrace{e^{-\frac{1}{2}\{\sigma^{BA}(12)+\sigma^{BA}(20)+\sigma^{BA}(10)\}}
-e^{-\frac{1}{2}\{\sigma^{BA}(12)+\sigma^{BA}(20)+\sigma^{BA}(10)\}
+\hat\sigma_{in}(12|10)}}\vspace{1cm} \nonumber\\
&& =1-e^{-\frac{1}{2}\{\sigma^{BA}(12)+\sigma^{BA}(20)\}}-e^{-\frac{1}{2}\sigma^{BA}(10)}
-e^{-\frac{1}{2}\{\sigma^{BA}(12)+\sigma^{BA}(20)+\sigma^{BA}(10)\}
+\hat\sigma_{in}(12|10)} \nonumber
\end{eqnarray}
\normalsize

All other splittings for type $B$ diagrams are found in a similar way by replacing
$\hat\sigma_{in}(12|10)$ by the proper inelastic term, same for $B^*$.
The term resulting from diagram $D$ in \fig{fig:real-virt} is readily calculated as

 \begin{eqnarray}\label{glaubBinel1}
\small
&& \hspace{2cm}\mbox{\tiny elastic scattering in the amplitude and conjugated amplitude}
\nonumber\\
&& \overbrace{\left(1-e^{-\frac{1}{2}\sigma^{BA}(10)}\right)
\left(1-e^{-\frac{1}{2}\sigma^{BA}(10)}\right)
}
\hspace{1cm}
\\
&&\hspace{2cm}\mbox{\tiny inelastic scattering in the amplitude and conjugated amplitude}  \nonumber\\
&&+
\overbrace{e^{-\frac{1}{2}2\sigma^{BA}(10)}
-e^{-\frac{1}{2}2\sigma^{BA}(10)
+\hat\sigma_{in}(10|10)}}\vspace{1cm} \nonumber\\
&& =1-e^{-\frac{1}{2}\sigma^{BA}(10)}-e^{-\frac{1}{2}\sigma^{BA}(10)}
-e^{-\frac{1}{2}2\sigma^{BA}(10)
+\hat\sigma_{in}(10|10)} \nonumber
\end{eqnarray}
\normalsize
The crossed dipole splittings $\frac{x_{12}}{x^2_{12}}\frac{x_{02'}}{x^2_{02'}}$
and $\frac{x_{02}}{x^2_{02}}\frac{x_{12'}}{x^2_{12'}}$ in $D$-type diagrams, has inelastic cross section suppressed in the large $N_C$ limit giving just the elastic term.
One can see that this way we reproduce the single inclusive cross section in the Glauber form
 Eqs.~(\ref{lineIN1})-(\ref{lineIN4})  found from a general expression Eqs.~(\ref{line1})-(\ref{line4}).

\end{document}